\documentclass[letterpaper,twocolumn,10pt]{article}
\usepackage{usenix}
\usepackage{balance} 

\usepackage{hyperref}
\usepackage[
	acronym,
	nomain,
	nopostdot,
	nogroupskip
]{glossaries}
\setacronymstyle{long-short}
\newacronym{ai}{AI}{artificial intelligence}
\newacronym{ran}{RAN}{radio access network}
\newacronym{gpu}{GPU}{graphics processing unit}
\newacronym{fm}{FM}{foundation model}
\newacronym{llm}{LLM}{large language model}
\newacronym{ml}{ML}{machine learning}

\newacronym{mac}{MAC}{medium access control}
\newacronym{nr}{NR}{new radio}
\newacronym{vran}{vRAN}{virtualized RAN}
\newacronym{oran}{O-RAN}{open RAN}
\newacronym{du}{DU}{distributed unit}
\newacronym{phy}{PHY}{physical layer}
\newacronym{fec}{FEC}{forward error correction}
\newacronym{ldpc}{LDPC}{low-density parity-check}
\newacronym{crc}{CRC}{cyclic redundancy check}
\newacronym{fft}{FFT}{fast Fourier transform}
\newacronym{ue}{UE}{user equipment}
\newacronym{prb}{PRB}{physical resource block}
\newacronym{mcs}{MCS}{modulation and coding scheme}
\newacronym{snr}{SNR}{signal-to-noise ratio}
\newacronym{tb}{TB}{transport block}
\newacronym{harq}{HARQ}{hybrid automatic repeat request}
\newacronym{ul}{UL}{uplink}
\newacronym{dl}{DL}{downlink}
\newacronym{ota}{OTA}{over-the-air}
\newacronym{scs}{SCS}{subcarrier spacing}
\newacronym{sa}{SA}{standalone}
\newacronym{ha}{HA}{hardware accelerator}

\newacronym{sm}{SM}{streaming multiprocessor}
\newacronym{smu}{SMU}{SM utilization}
\newacronym{acu}{ACU}{arithmetic compute utilization}
\newacronym{gbu}{GBU}{global bandwidth utilization}
\newacronym{gemm}{GEMM}{general matrix multiplication}
\newacronym{mps}{MPS}{multi-process service}
\newacronym{mig}{MIG}{multi-instance GPU}

\newacronym{dag}{DAG}{directed acyclic graph}
\newacronym{dp}{DP}{data parallelism}
\newacronym{pp}{PP}{pipeline parallelism}
\newacronym{sgd}{SGD}{stochastic gradient descent}
\newacronym{air}{AIR}{asynchronous inter-site rerouting}
\newacronym{mlr}{MLR}{model layout rebalancing}
\newacronym{cas}{CAS}{compare-and-swap}
\newacronym{ewma}{EWMA}{exponentially weighted moving average}
\newacronym{cov}{CoV}{coefficient of variation}

\glsdisablehyper

\usepackage{amsmath}
\usepackage{amssymb}
\usepackage{multirow}
\usepackage{multicol}
\usepackage[most]{tcolorbox}
\usepackage{dsfont}
\usepackage{enumitem}

\usepackage{subcaption}
\usepackage{graphicx}
\usepackage{wrapfig}
\usepackage{algorithm}
\usepackage{algorithmic}
\usepackage{booktabs}

\usepackage{tikz}
\usetikzlibrary{shapes.geometric}
\usetikzlibrary{arrows.meta,positioning}

\usepackage{cleveref}
\usepackage{xspace}
\usepackage{comment}
\newcommand{\sys}{\textsc{Weaver}\xspace}

\newcommand{\tinyskip}{}
\newcommand{\mypar}[1]{\tinyskip\noindent\textbf{#1.}\xspace}

\usepackage{comment}

\usepackage{mathtools}
\newcommand{\eg}{\text{e.g.,}\ }

\newenvironment{tightlist}{
	\begin{list}{$\bullet$}{
			\setlength{\topsep}{.1em}
			\setlength{\partopsep}{0in}
			\setlength{\parskip}{0in}
			\setlength{\itemsep}{0in}
			\setlength{\parsep}{0in}
			\setlength{\leftmargin}{1em}
			\setlength{\rightmargin}{0in}
			\setlength{\itemindent}{0in}
		}}
		{\end{list}}

\newcommand{\markReshard}{\textcolor{orange}{$\bigstar$}}
\renewcommand{\thefootnote}{\arabic{footnote}}
\newcommand{\algorithmstyle}[1]{\renewcommand{\algocf@style}{#1}}

\newlength\myindent
\usepackage{siunitx}
\renewcommand{\sectionautorefname}{\S{}}

\crefname{section}{§}{§§}
\crefname{section}{§}{§§}
\crefformat{section}{§#2#1#3}
\crefname{table}{Table}{Table}
\crefname{figure}{Fig.}{Fig.}
\crefname{listing}{Listing}{Listing}
\crefname{algorithm}{Alg.}{Alg.}
\crefname{objective}{\sc{\textbf{Objective}}}{\sc{\textbf{Objective}}}
\newcommand{\floatfont}{\footnotesize}
\newcommand{\clampbox}[2]{{\floatfont\resizebox{\ifdim\width>#1 #1\else\width\fi}{!}{#2}}}

\begin{document}
\title{\sys: A System for AI-RAN Compute Sharing with Foundation Model Training}

\author{
Leyang Xue$^{1*}$, 
Tianxin Wang$^{1*}$,
Xin Zhe Khooi$^{2*}$, 
Jiaxun Yang$^{1}$,
Dheeraj Mahendiran$^{1}$, \\
Yufeng Xia$^{1}$, 
Mun Choon Chan$^{2}$,
Myungjin Lee$^{3}$,
Mahesh K. Marina$^{1}$ \\[0.5em]
{\small\normalfont $^{1}$The University of Edinburgh\quad
$^{2}$National University of Singapore\quad
$^{3}$Cisco Research}
}

\maketitle
\begingroup
\renewcommand{\thefootnote}{\fnsymbol{footnote}}
\footnotetext[1]{These authors contributed equally to this work.}
\endgroup

\maketitle

\begin{abstract}
The emergence of AI-RAN infrastructure, which equips cell sites with GPU-accelerated hardware, creates an opportunity to colocate non-RAN workloads with primary RAN processing. 
We explore using this spare capacity for decentralized training of foundation models (FMs), one of the most compute-intensive AI workloads. 
We present the first characterization of spare GPU capacity in AI-RAN systems at both micro-scale—across transmission slots within a cell site—and macro-scale—across sites. 
Our analysis finds that 40–85\% of GPU capacity is unused; although this capacity is temporally bursty at individual sites, it is spatially complementary across sites. 
To safely and efficiently harness these resources, we present \sys, a system that opportunistically trains FMs alongside latency-critical RAN workloads without degrading RAN performance. 
\sys adopts a RAN-first design: a spare-compute controller integrated into the MAC scheduler uses compute-aware scheduling to smooth RAN GPU demand and exposes more usable spare GPU capacity. 
A two-level elastic training framework then adapts to dynamic, heterogeneous spare capacity within and across sites. 
Experiments on an O-RAN-aligned system prototype show that \sys creates up to 4.9$\times$ more usable spare compute and utilizes up to 83\% of the available spare capacity. 
On a multi-site testbed, \sys improves training throughput by 2.1–3.7$\times$ over baseline approaches.
\end{abstract}

\section{Introduction}

As we head towards 6G, there is a wide recognition that AI for radio access network (RAN) -- ``AI-for-RAN'' -- can unlock significant spectral, energy, and operational efficiency gains (e.g., \cite{shafin-mwc20,spotlight,kairos,symbxrl,iridescence}).
Building on this, there is growing momentum towards a more holistic integration of AI with the RAN to realize AI-RAN~\cite{ai-ran}, which goes beyond AI-for-RAN to also cover ``AI-on-RAN'' to enable novel edge (AI) services~\cite{tutti,jin-conext25}, as well as ``AI-and-RAN'' to share RAN compute infrastructure between RAN and non-RAN (AI) workloads~\cite{yinYangRAN}.
As envisioned by the AI-RAN Alliance~\cite{ai-ran-alliance}, AI-RAN promises not only reduced operational costs through various efficiency gains but also helps maximize RAN asset utilization and generate new revenue streams (e.g., through value-added services like sensing).
Enabling the full potential of AI-RAN does entail upgrading existing mobile network infrastructure at cell sites to embed accelerators with RAN compute hardware that can natively support AI workloads.
Recent developments, including the emergence of NVIDIA AI Aerial platforms~\cite{nvidia-aerial-platforms} and Nokia's partnership with NVIDIA~\cite{nokia-nvidia}, indicate movement in this direction.

We believe the shift toward GPU-accelerated RAN infrastructure for AI-native 6G networks presents a compelling and timely opportunity for hosting non-RAN AI workloads on AI-RAN compute infrastructure, aligned with the AI-and-RAN aspect of AI-RAN. 
Looking ahead, potential AI compute distributed across the RAN infrastructure can be enormous. A back-of-the-envelope calculation suggests that the aggregate GPU compute capacity across RAN cell sites could potentially rival the currently deployed global GPU compute capacity.\footnote{There are currently around 20 million 4G/5G RAN cells globally~\cite{opencellid}. Assuming the number of cells remains at least at that level going forward to 6G, deploying a GH200 server (each with about 1 PFLOP capacity) for every 20 cells, as per the benchmarking in \cite{nvidia-gh200-ran}, yields an estimated total of 1 ZFLOP across all RAN sites. Current deployed GPU capacity globally is around 4 ZFLOPS~\cite{epoch-gpu-capacity}.}
Not only that, such compute would typically be only partially utilized for RAN processing. RAN compute usage varies with the traffic it carries while compute infrastructure is typically provisioned for the peak usage (worst case). The spatiotemporal variations inherent to RAN traffic at both micro and macro scales~\cite{xu-tnet16,ShanghaiDataset,SpectraGAN,concordia,yinYangRAN} may therefore create plentiful spare compute that can be harnessed by other AI workloads.

In this paper, we explore the aforementioned opportunity considering foundation model (FM) training as a representative yet challenging non-RAN AI workload.
FMs~\cite{foundation-models} are models trained at scale on broad data that can then be adapted to a wide range of downstream tasks (e.g., GPT-4~\cite{gpt4}, Stable Diffusion~\cite{stable-diffusion}, AlphaFold~\cite{alphafold}).
\Glspl{llm} such as GPT-4 are a prominent subclass of foundation models that are trained on language data.
FM training requires enormous compute.
For example with LLMs, scaling model size, data size, and training duration consistently improves capability, driving rapid growth in computational demand~\cite{scalelaw}. 
Growing interest in domain-specific FMs, including for mobile networks~\cite{telecom-fm}, further amplifies this demand.
This has in turn contributed to a highly \emph{centralized, cloud-centric AI ecosystem} currently.
In view of the above, leveraging the spare compute on naturally decentralized AI-RAN compute infrastructure can serve as an alternative platform for FM training that is complementary to the cloud. 

FM training over AI-RAN compute infrastructure, however, requires addressing \textbf{three main challenges}:
\begin{enumerate}[leftmargin=*,nosep] 
    \item We need to clearly understand the scale and nature of spare compute likely to be available on an AI-RAN compute (GPU-accelerated RAN) infrastructure.
    Prior work offers limited insight, both in the scope of metrics and spatiotemporal scales considered, as discussed in \S\ref{sec:spare-compute} and \S\ref{sec:related-work}.
    
    \item Even when spare compute exists, making it usable for non-RAN workloads such as FM training is challenging.
    This is because RAN traffic is inherently bursty at microscopic scales~\cite{concordia}, which in turn manifests as burstiness in RAN related GPU compute usage.
    Smoothing that burstiness to maximize usable spare GPU compute must be done carefully to ensure RAN performance is unaffected.
    \item Efficiently utilizing the available spare GPU compute resource for FM training within and across cell sites in an AI-RAN compute infrastructure necessitates a training strategy that is robust and efficient in the face of resource dynamism and heterogeneity across sites and timescales.
\end{enumerate}

To tackle the first challenge, we conduct the first comprehensive characterization study on spare GPU compute capacity in GPU-accelerated RAN infrastructure, both at the micro-scale (RAN-slot-level) within a cell site and at the macro-scale across cell sites (\S\ref{sec:spare-compute}), focusing on offloading computationally heavy part of RAN processing on LDPC decoding to the GPU as in prior work~\cite{CloudRIC,yinYangRAN,decodeX}.
Our study considers NVIDIA DGX Spark as a representative AI-RAN compute platform along with large-scale production mobile network traffic trace.
Our study reveals that substantial unused GPU resources (overall, about 40--85\%) exist that could be harnessed by non-RAN workloads.
We observe that memory bandwidth-limited RAN processing complements compute-bound FM training. 
At a macro scale, spare compute mirrors diurnal traffic patterns at individual cell sites, while being spatially diverse across sites. 

Building on the above outlined spare-compute characterization, we address the second and third challenges with a novel RAN-first system architecture design termed \sys (\S\ref{sec:system-overview}), illustrated in \cref{fig:overview}. 
\sys enables efficient FM training on GPU-accelerated RAN infrastructure by opportunistically sharing GPU resources with RAN processing while protecting RAN performance.  
\sys employs Green Contexts based dynamic GPU sharing (\cref{tab:gpu-sharing}).
The \sys system is made up of two principal components:
\begin{tightlist}
\item First, \sys introduces a \emph{RAN-centric spare compute controller} (\S\ref{sec:ran-control}) embedded inside the RAN stack, in contrast to relying on externally predicting RAN-related GPU compute demand as done by prior works~\cite{yinYangRAN}.
By integrating with the MAC scheduler, it performs compute-aware scheduling to smooth RAN related GPU usage (despite RAN traffic burstiness) to unlock maximal usable spare GPU resource for non-RAN workloads while enforcing per-slot deadlines to guarantee RAN performance. 
\item Second, \sys introduces a tailored \emph{two-level elastic FM training framework} (\S\ref{sec:training-strategy}) to efficiently cope with spatiotemporal variations in spare GPU compute within and across cell sites. 
At the inter-site level, a centralized coordinator dynamically redistributes training computation and selectively reconfigures model layouts across cell sites only when long-term compute/traffic shifts occur.
Within a cell site (i.e., intra-site), \gls{gemm} operations (that dominate FM training) are split into fine-grained and elastic tiles to match the instantaneous available spare compute.
\end{tightlist}

A publish–subscribe interface couples these two components, allowing FM training to continuously adapt to the RAN-approved spare GPU resource at each cell site.
Together, these design choices enable safe, efficient co-location of latency-critical RAN processing and opportunistic FM training.    

We conduct a comprehensive evaluation of \sys spanning both its main components (\S\ref{sec:evaluation}). 
We build a system prototype that deploys \sys{}’s RAN-centric spare-compute controller on an O-RAN-aligned 5G NR stack with GPU-accelerated LDPC decoding~\cite{openairinterface-gpu}, while co-locating FM training computation on the same GPU.
Results from evaluating the system under production traffic traces and live end-to-end UE transmissions show that \sys{}'s RAN-centric approach reduces fluctuations in RAN related GPU usage by up to 4.9$\times$ without degrading RAN performance, while exposing substantially more usable spare compute to the co-located training workload than prior and baseline methods. 
Then, on an eight-site distributed GPU testbed, \sys{}’s two-level elastic FM training framework achieves 2.1--3.7$\times$ higher training throughput than baseline systems while harvesting up to 83\% of available spare compute. 
Finally, via simulations, we show that \sys{}'s training framework effectively scales to hundreds of sites and larger models.

The following section provides the essential background before we proceed to describing the main contributions of the paper in the subsequent sections.

\section{Background}
\label{sec:background}

\mypar{GPU acceleration for vRAN} 
Virtualized Radio Access Network (vRAN) moves baseband processing onto general-purpose platforms~\cite{bonati-comnet20,concordia}. 
The Distributed Unit (DU) hosts the most compute-intensive MAC and PHY functions, with LDPC decoding as the dominant bottleneck~\cite{openairinterface-gpu,CloudRIC,2022-MEMU-5GPerf}. 
Because LDPC decoding must meet strict millisecond-scale deadlines while maintaining carrier-grade reliability, hardware acceleration is often required in production deployments~\cite{openairinterface-gpu}. 
While ASICs and FPGAs have traditionally served this role, GPUs are increasingly attractive due to their massive parallelism and programmability~\cite{yinYangRAN}. 
For example, NVIDIA's SionnaRK~\cite{openairinterface-gpu} accelerates LDPC decoding, while Aerial~\cite{nvidia-aerial,nvidia-aerial-cuphy} offloads the entire 5G PHY pipeline to GPUs, demonstrating their viability for compute-intensive vRAN processing.

\mypar{GPU architecture overview}
A modern NVIDIA-style GPU consists of tens to hundreds of \emph{Streaming Multiprocessors (SMs)}, each hosting hardware threads and execution pipelines, while all SMs share a high-bandwidth global memory~\cite{cuda-prog-guide}.
The GPU programming model exposes these resources via \emph{kernels}, which are functions launched by the host CPU, each with a specified SM demand.

\mypar{GPU utilization metrics}
GPU load can be characterized using three metrics:
(i)~\emph{\Glsentrylong{smu} (\glsentryshort{smu})}\glsunset{smu}, the fraction of SMs assigned to a kernel;
(ii)~\emph{\Glsentrylong{acu} (\glsentryshort{acu})}\glsunset{acu}, the fraction of peak sustained arithmetic issue rate achieved;
and (iii)~\emph{\Glsentrylong{gbu} (\glsentryshort{gbu})}\glsunset{gbu}, the fraction of peak global-memory bandwidth achieved.

\mypar{GPU sharing for AI-and-RAN} 
Prior work has explored GPU sharing for co-locating RAN and AI workloads. 
YinYangRAN~\cite{yinYangRAN} uses \gls{mps} to partition SMs between 5G DU PHY processing and ML inference, while CAORA~\cite{CAORA} employs \gls{mig} to create isolated GPU instances. 
However, both approaches rely on RAN workload prediction for dynamic resource provisioning. 
Mis-estimation is problematic: overestimation starves co-located workloads, whereas underestimation may violate PHY processing deadlines. 
They also incur substantial reconfiguration overheads (${ \sim }$0.3\,s for MPS and ${ \sim }$7\,s for MIG), during which the GPU cannot service workloads or processing must fall back to CPUs, making them impractical for realtime-sensitive and mission-critical RAN workloads.

More broadly, modern GPUs expose a range of temporal, spatial, and hybrid sharing mechanisms~\cite{cuda-prog-guide} (see \cref{tab:gpu-sharing}). 
However, existing GPU-sharing systems are designed for multi-tenant environments where workloads are treated as peers~\cite{gpreempt,krypton,lithos,pantheon}. 
In contrast, RAN workloads must be unconditionally prioritized and require per-slot ($\leq$1\,ms) deadline guarantees. 
While some systems distinguish latency-sensitive and best-effort tenants~\cite{orion,reef,xsched}, they do not provide the hard isolation and deadline guarantees required for RAN processing. 
Among existing mechanisms, \emph{Green Contexts} are unique in combining hardware SM isolation with microsecond-scale reconfiguration, making them the only practical substrate for per-slot RAN control, which we will later explore in this paper. 
Moreover, Green Contexts are natively supported in the NVIDIA AI-RAN software stack~\cite{nvidia_aerial_cuda_accelerated_ran}. 

\begin{table}[t]
	\centering
	\caption{Types of GPU sharing in current systems.}
	\label{tab:gpu-sharing}
	\clampbox{\linewidth}{
		\setlength{\tabcolsep}{3pt}
		\begin{tabular}{rlllr}
			\toprule
			\textbf{Paradigm}         & \textbf{Mechanism}                        & \textbf{Isolation}              & \multicolumn{2}{c}{\textbf{Reconfiguration}}         \\
			\midrule
			\multirow{2}{*}{Temporal} & Context switch~\cite{gpreempt}            & Device exclusive                & Restore              & $\sim$100\,ms                 \\
			                          & Preemption~\cite{reef,xsched}             & Instruction-level               & Application rebuild  & --                            \\
			\midrule
			\multirow{2}{*}{Spatial}  & MIG~\cite{krypton}                        & SM \& memory                    & GPU reset            & $\sim$7\,s                    \\
			                          & \textbf{Green Contexts}~\cite{cuda-prog-guide} & \textbf{SM}                & \textbf{CUDA Stream} & {\bfseries\boldmath$\sim$10\,$\mu$s} \\
			\midrule
			\multirow{2}{*}{Mixed}    & MPS~\cite{CloudRIC}                       & Best-effort SM (no mem.\ isol.) & GPU reset            & $\sim$300\,ms                 \\
			                          & CUDA streams~\cite{orion,cuda-prog-guide} & Best effort                     & Static               & $\sim$1\,$\mu$s               \\
			\bottomrule
		\end{tabular}
	}
\end{table}

\mypar{Foundation model training} 
FM training involves distributed computation over a \gls{dag} of dense matrix multiplications (\glspl{gemm})~\cite{palm,megatron}. 
Each layer produces a sequence of \gls{gemm} operations arising from attention projections, feed-forward transforms, and embedding lookups, whose execution order and synchronization points are determined by the distributed training strategy used (\eg \gls{dp}, \gls{pp}, or their combination). 
Over 99\,\% of training FLOPs are \glspl{gemm}~\cite{palm,megatron}, making FM training compute-bound and capable of saturating the GPU's arithmetic pipelines rather than its memory bus. 
We leverage this complementary resource profile (see later in~\cref{sec:micro-scale}) to co-locate FM training with memory-bound RAN workloads and exploit otherwise idle compute capacity.

\section{Spare Compute Characterization}
\label{sec:spare-compute}

The adoption of GPUs in vRAN raises a fundamental question: 
\emph{how much spare compute is available, and along which dimensions can it be safely exploited without affecting RAN performance?} 
We study this at two levels: 
(i) micro-scale (\cref{sec:micro-scale}), examining fine-grained opportunities within slots and across GPU resources (compute, memory bandwidth, and SM occupancy) across a diverse range of RAN workloads; and 
(ii) macro-scale (\cref{sec:rq1-macro-spatial}), analyzing whether spare capacity remains consistently available over time and across cell sites.

Answering this question requires more than the coarse scalar utilization metrics reported in prior work (e.g., CloudRIC~\cite{CloudRIC}); see \S\ref{sec:related-work}. 
At the micro-scale, standard tools (e.g., \texttt{nvidia-smi}) provide only coarse activity indicators and do not expose multi-dimensional GPU utilization~\cite{dcgm}. 
At the macro-scale, traffic traces~\cite{imdea} capture demand dynamics but cannot directly reveal actual GPU utilization.

In this work, our focus is on vRAN deployments with GPU-accelerated LDPC decoding, inline with prior work~\cite{CloudRIC,yinYangRAN,decodeX}, given that it is a widely known compute bottleneck~\cite{2022-MEMU-5GPerf}.

\subsection{Micro-Scale Opportunities}
\label{sec:micro-scale}

\begin{figure}[t]
	\centering
	\includegraphics[width=\linewidth]{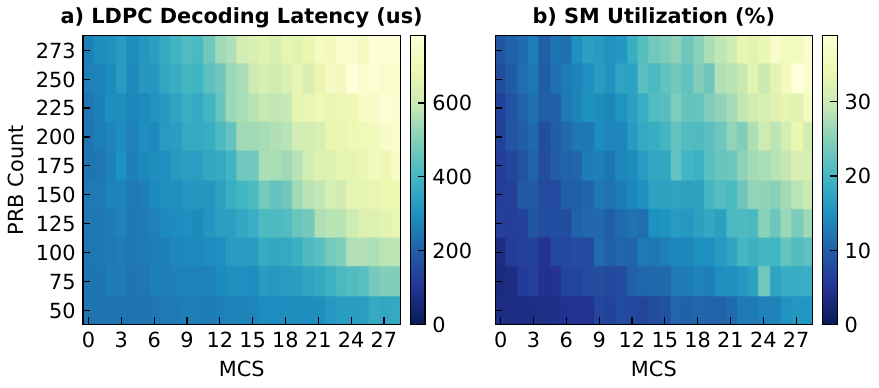}
	\includegraphics[width=\linewidth]{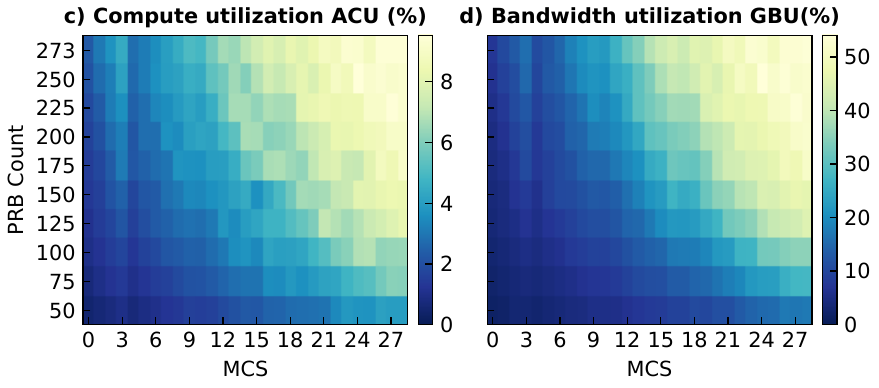}
	\caption{a) Worst-case latency, b) SMU, c) ACU, and d) GBU of LDPC decoding across different \#PRBs and MCS.}
	\label{fig:ldpc_combined}
\end{figure}

First, to study the micro-scale opportunities across slots within a cell, we use the OpenAirInterface5G-based SionnaRK~\cite{openairinterface-gpu} to profile GPU-accelerated LDPC decoding on an NVIDIA DGX Spark (48 SMs). 
We collect measurements using \texttt{ulsim} to sweep across the full modulation and coding scheme (MCS 0--28) and physical resource block (PRB 0--273) configuration space. 
To capture worst-case execution time, the LDPC decoder is fixed at 10 iterations.

\mypar{Slack within a slot}
We observe two sources of spare GPU capacity within a slot. 
First, even the worst-case RAN configuration completes LDPC decoding in $\sim$700$\mu${s}, leaving 30--70\% of the headroom available given a typical 1-ms deadline~\cite{CloudRIC} (\cref{fig:ldpc_combined}a). 
Second, GPU resources are underutilized: worst-case SM utilization reaches only $\sim$35\%, for a single DU, and remains below 20\% across most operating points (MCS $\le$ 20, PRB $\le$ 200), leaving over 80\% of the available SMs on the GPU idle. 
Thus, LDPC decoding primarily scales in execution time rather than SM occupancy (\cref{fig:ldpc_combined}b).

\begin{figure}[t]
	\centering
	\includegraphics[width=\linewidth]{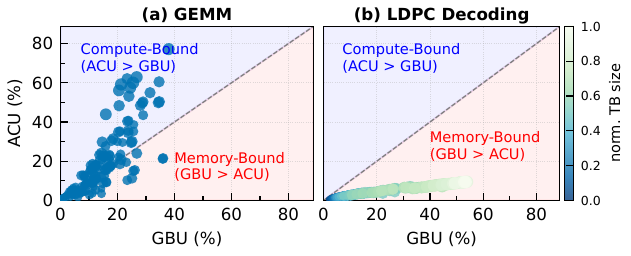}
	\caption{ACU and GBU of GPU operations for FM training (model sizes: 1.3B to 70B) vs. LDPC decoding.}
	\label{fig:gemm}
\end{figure}

\mypar{Microscopic (compute and memory) view of slack} 
Examining spare GPU resources after RAN execution (i.e., LDPC decoding) reveals two key observations. 
First, RAN is not compute intensive: even under the worst-case configuration, the Arithmetic Compute Utilization (\gls{acu}) reaches only $\sim$12\% (\cref{fig:ldpc_combined}c). 
This stems from the latency-driven nature of RAN processing: LDPC decoding is parallelized across SMs to minimize completion time, resulting in low ACU.
Second, RAN is memory-bandwidth intensive: Global Bandwidth Utilization (\gls{gbu}) ranges from 10--60\% depending on transport block (TB) size and redundancy (\cref{fig:ldpc_combined}d), indicating that LDPC decoding is memory bandwidth-bound (see detailed breakdown in Appendix~\ref{app:ldpc-intensity}) while leaving much of the compute pipeline idle.
This creates a natural complementarity with FM training, which is compute-intensive (see~\cref{fig:gemm}). 

\subsection{Macro-Scale Opportunities}
\label{sec:rq1-macro-spatial}

Next, to assess whether spare capacity remains consistently available over time and across sites, we replay a large-scale production traffic trace from six urban and suburban cell sites~\cite{imdea} using a modified version of the OpenAirInterface (OAI) \texttt{phy-test} tool~\cite{mani2025oai5gran}. 
For brevity, we report only SMU, as ACU and GBU exhibit similar trends (\cref{sec:micro-scale}).

\begin{figure}[t]
	\includegraphics[width=\linewidth]{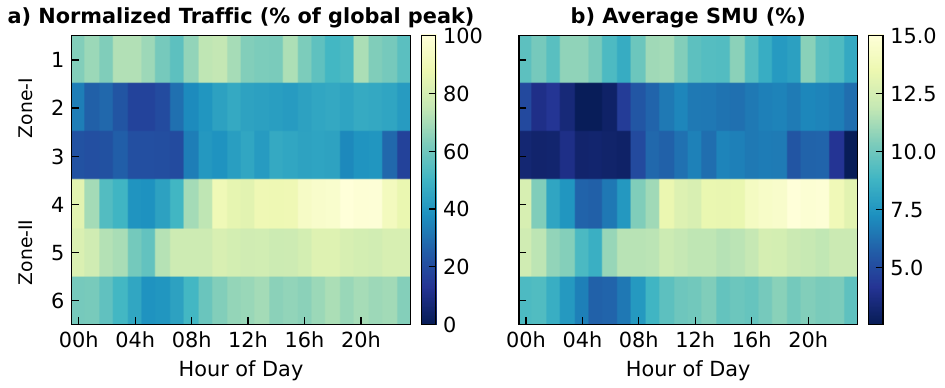}
	\caption{Spatial traffic patterns and their SMU.}
	\label{fig:spare-compute-summary}
\end{figure}

\mypar{Single-site temporal spare-compute patterns} 
We observe a clear diurnal pattern in both traffic and SMU across sites. 
Traffic is generally low and stable overnight (02:00--06:00) and peaks during the day (\cref{fig:spare-compute-summary}a).
Correspondingly, the available spare compute is highest overnight, while even at the busiest site during peak demand, roughly 85\% of SM capacity remains available (\cref{fig:spare-compute-summary}b).

\mypar{Spatial cross-site spare-compute patterns}
We compare traffic demand and SMU across six sites, revealing three observations.
First, \emph{highly skewed traffic}:  Zone-II (suburban) sites (4--6) carry noticeably higher load than Zone-I (urban) sites (1--3), leaving low-traffic sites with significant spare compute even during local peaks (\cref{fig:spare-compute-summary}a). 
Second, \emph{staggered daily peaks}: sites peak at different times of day, so simultaneous peak load across sites is rare (\cref{fig:spare-compute-summary}a). 
Third, \emph{SMU tracks traffic}: SMU follows the same spatiotemporal pattern as traffic, showing that spare compute is abundant at individual sites and complementary across sites (\cref{fig:spare-compute-summary}b).

\begin{tcolorbox} [colback=gray!10, colframe=gray!40, boxrule=0.3pt, arc=1mm, left=1mm,right=1mm,top=0.5mm,bottom=0.5mm, enhanced, before skip=2mm, after skip=2mm] 
\textbf{Key takeaway:}
Spare GPU capacity exists at multiple scales. 
Microscopically, LDPC decoding is memory-bound (complementary to compute-bound FM training) and leaves substantial compute resources idle. 
Macroscopically, this slack persists across time and space (cell sites), making co-location with FM training feasible.
\end{tcolorbox}

\section{\sys System Overview}
\label{sec:system-overview}

Building on the characterization study in \cref{sec:spare-compute}, \sys{} seeks to exploit the substantial spare GPU capacity available in vRAN deployments for FM training. 
Driven by that target, \sys{} is designed with the following goals:

\begin{enumerate}[leftmargin=*,nosep] 
    \item 
    \textbf{Maximize harvestable spare compute.} 
    \sys{} must expose as much GPU slack as possible to FM training without compromising RAN performance. 
    
    \item 
    \textbf{Efficiently utilize distributed slack.} 
    \sys{} must allow FM training to effectively exploit spare compute distributed across time and cell sites. 
\end{enumerate}

\begin{figure}[t]
	\centering
	\includegraphics[width=\linewidth]{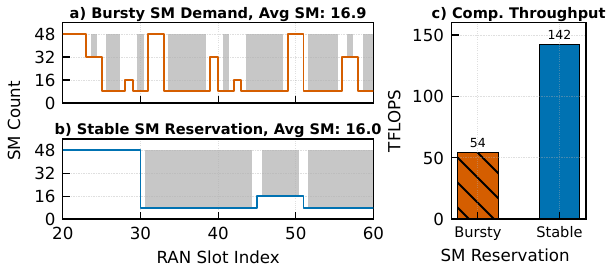}
	\caption{SM reservation comparison: a) bursty SM demand derived directly from RAN traffic; b) stable SM reservation; c) impact of SM reservation stability on training computational throughput (the higher, the better).}
	\label{fig:sm-impact-on-training}
\end{figure}

\mypar{Challenges and solutions}
Achieving these goals is challenging for three reasons. 
Below, we describe each challenge and the corresponding solution adopted in \sys{}.

\begin{enumerate}[leftmargin=*, nosep]
    \item 
    \textbf{Only SM allocation is controllable.} 
    Our characterization reveals spare capacity across multiple dimensions, including SMU, ACU, and GBU (\cref{sec:micro-scale}). 
    However, current GPUs expose only SM allocation as a practical runtime control knob~\cite{cuda-prog-guide}; ACU and GBU cannot be directly partitioned or allocated. 
    Consequently, any sharing mechanism must expose spare compute through SM allocation; Appendix~\ref{app:sm-enveloping-effect} shows that capping the RAN's SM count also frees headroom in all three dimensions. 
    Among available GPU-sharing primitives (\cref{sec:background}), Green Contexts are uniquely suited for this purpose, providing hardware-isolated SM partitions with microsecond-scale reconfiguration. 
    We therefore define an \textit{SM reservation}: the per-slot SM allocation reserved for RAN processing, whose complement is made available to FM training.

    \item 
    \textbf{The SM reservation is highly dynamic.} 
    The amount of SMs required by RAN processing is not constant.
    Bursty RAN traffic~\cite{imdea} causes frequent fluctuations in SM demand (\cref{fig:sm-impact-on-training}a). 
    While Green Contexts enable fine-grained SM partitioning, exposing these fluctuations directly to FM training results in a rapidly changing compute budget. 
    As shown in \cref{fig:sm-impact-on-training}, such variability reduces the effective computational capacity available to GEMM workloads, both through frequent SM reconfigurations and the disruption of long-running training kernels. 
    The challenge therefore is to transform a highly dynamic SM demand from the RAN into a stable spare-compute budget without compromising RAN correctness. 
    To address this, \sys{} introduces a \emph{RAN-centric spare compute controller} (\cref{sec:ran-control}) that derives a stable SM reservation for RAN to maximize spare compute and minimize reconfiguration overhead, while preserving sufficient headroom for RAN processing to absorb short-term fluctuations in RAN demand.
    
    \item 
    \textbf{Spare compute is distributed and time-varying.} 
    Available slack varies across both time and cell sites (\cref{sec:rq1-macro-spatial}), creating dynamic and heterogeneous available GPU compute across training workers. 
    Without adaptation, stragglers would limit distributed-training throughput. 
    To address this, \sys{} introduces \emph{two-level elastic training} (\cref{sec:training-strategy}), which elastically adjusts training execution both within a site and across sites to match available spare compute.
    
\end{enumerate}

\begin{figure}[t]
	\centering
	\includegraphics[width=\linewidth]{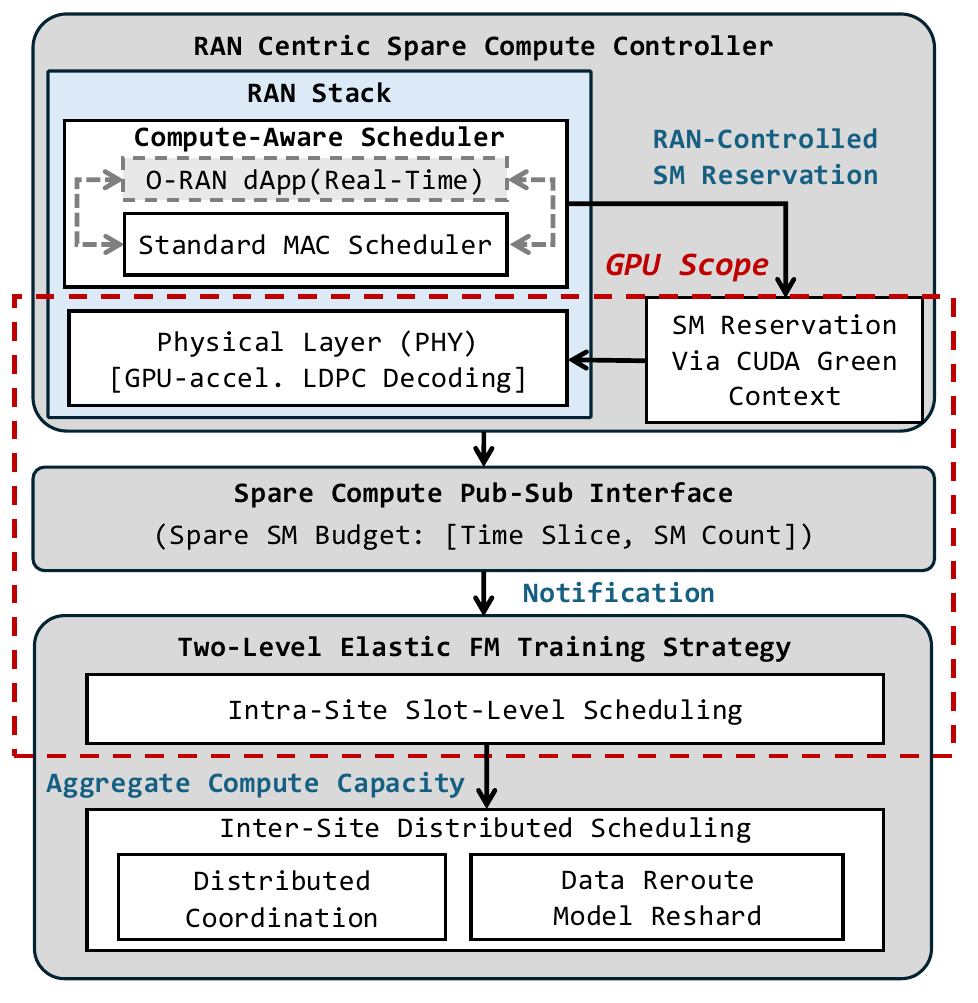}
	\caption{\sys{} system architecture overview.}
	\label{fig:overview}
\end{figure}

Taken together, these challenges motivate the \sys{} system architecture shown in~\cref{fig:overview}.
\sys{} combines (i) Green-Context-based SM reservation, (ii) a RAN-centric spare-compute controller that stabilizes the SM reservation to maximize usable spare compute, and (iii) a two-level elastic training framework that efficiently utilizes distributed spare compute. In \sys{}, a publish-subscribe (pub-sub) interface is used for (ii) to continually communicate RAN's SM reservation to (iii).
We describe the design of components (ii) and (iii) in detail in the following two sections.

\section{RAN-Centric Spare Compute Controller}
\label{sec:ran-control}

As discussed in \cref{sec:system-overview}, a key challenge is to transform highly dynamic RAN SM demand into a stable spare-compute budget that can be exploited for FM training. 
Prior approaches for AI-and-RAN infer available GPU resources \emph{externally} through RAN workload prediction~\cite{yinYangRAN}. 
However, RAN SM demand is highly dynamic, and so mis-prediction either risks RAN deadline violations or leaves spare compute underutilized. 
So, instead of predicting future demand, \sys{} adopts a \emph{RAN-first approach}.
Specifically, it introduces a RAN-centric spare-compute controller embedded inside the RAN stack that directly shapes RAN compute demand and exposes the remaining spare capacity to FM training -- a non-RAN AI workload we focus on. 

The MAC scheduler is the natural control point, as its scheduling decisions determine the amount of LDPC decoding work and thus the required SM reservation. 
The key idea behind the spare-compute controller in \sys{} is \emph{compute-aware scheduling}, which smooths the short-term fluctuations in RAN compute demand while preserving RAN processing deadline and QoS guarantees.
Rather than serving every burst immediately, \sys{} absorbs transient traffic spikes in UE buffers and only increases the RAN's SM reservation when backlog growth approaches a QoS-aware threshold. 
This effectively rebalances PRB allocation across neighboring slots, transforming bursty slot-level demand into a more stable SM reservation. 
The resulting spare-compute budget is substantially more useful for FM training.

To achieve this, \sys{} implements a stack-agnostic O-RAN dApp~\cite{doro2022dapps,oran_dapp_architecture_2026} that wraps the standard-aligned MAC scheduler and augments it with compute awareness (\autoref{fig:overview}).
The resulting per-slot SM reservation is enforced through CUDA Green Contexts, which provide hardware-isolated SM partitions and microsecond-scale reconfiguration.

\subsection{Compute-Aware Scheduling Pipeline}
\label{sec:compute-aware-sched}

\begin{figure}[t]
	\centering
	\includegraphics[width=0.95\linewidth]{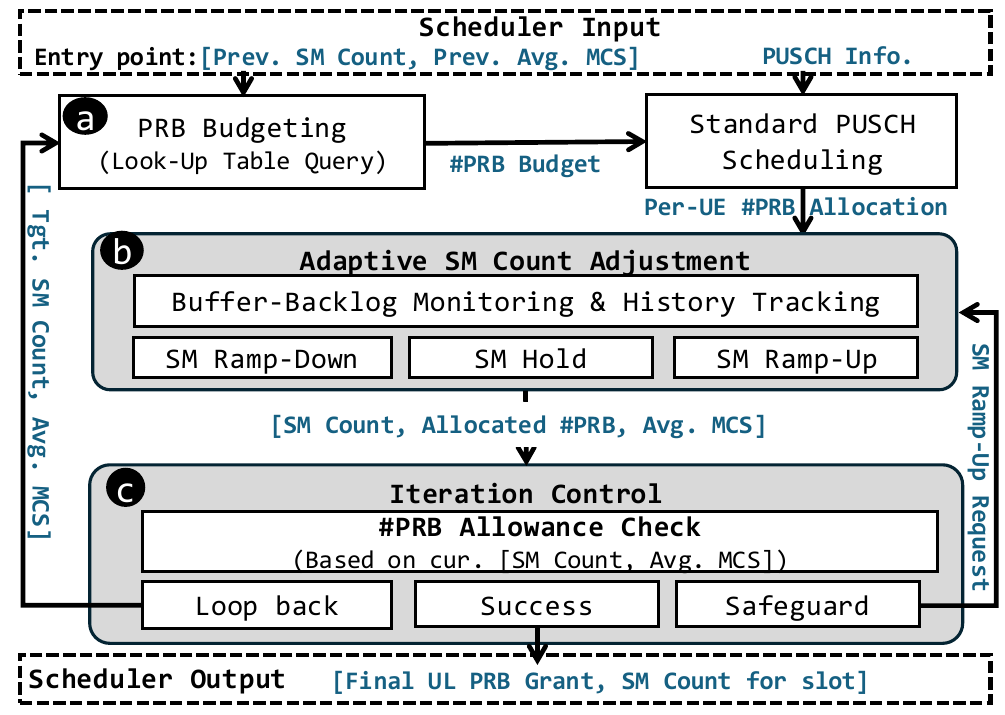}
	\caption{Per-slot compute-aware scheduling pipeline.}
	\label{fig:ran-control}
\end{figure}

\cref{fig:ran-control} depicts the compute-aware scheduling pipeline in \sys{}.
The pipeline comprises three key steps: 
\textcircled{a} PRB budgeting,
\textcircled{b} adaptive SM reservation adjustment, and
\textcircled{c} iteration control.
We elaborate on these steps below.

\subsubsection{PRB budgeting using lookup tables}

\begin{figure}[t]
	\centering
	\includegraphics[width=\linewidth]{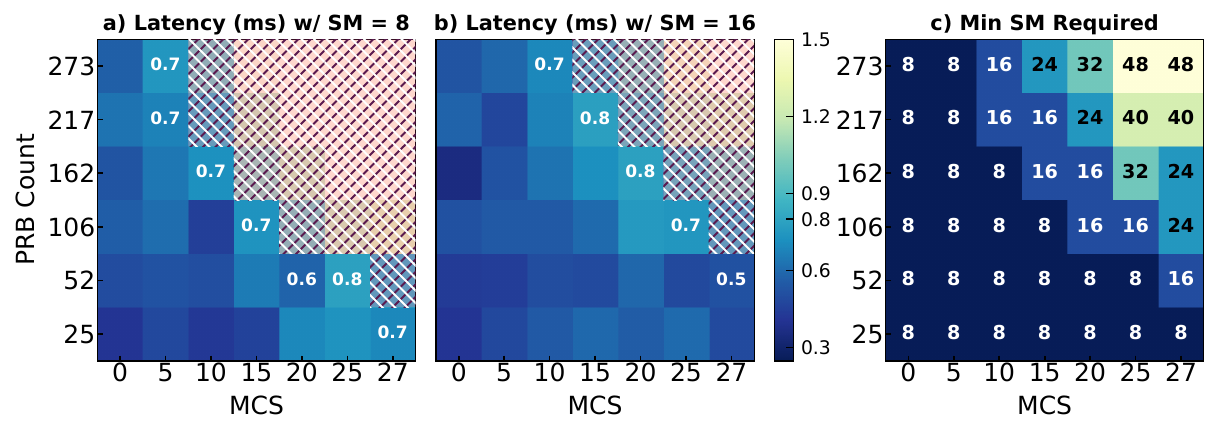}
	\caption{ a) and b): Worst-case LDPC decoding latency with various (PRB, MCS) tuples given an SM count; c): minimum SM demand as a function of (PRB, MCS).}
	\label{fig:sm-lookup-table}
\end{figure}

\mypar{Lookup table}
A key enabling property is that SM reservation for RAN is directly controllable through PRB allocation. 
Given an MCS index, capping the number of scheduled PRBs bounds the amount of LDPC decoding work and therefore the required SM count. 
We capture this relationship using an offline-generated lookup table (\cref{fig:sm-lookup-table}c), derived from worst-case LDPC decoding latency profiling across SM budgets and RAN configurations (e.g., \cref{fig:sm-lookup-table}a--b) using \texttt{ulsim} (\S\ref{sec:micro-scale}). 

As an example, (MCS~15, 162~PRBs) satisfies the latency deadline with 16~SMs but violates it with 8~SMs (\cref{fig:sm-lookup-table}b); the table therefore records 16 as the minimum required SM count (\cref{fig:sm-lookup-table}c). 
The resulting table provides safe operating bounds because it is derived from conservative worst-case profiling, covers the full MCS/PRB configuration space, and is calibrated through end-to-end testing under extreme channel and traffic conditions (Appendix~\ref{app:lookup-table}). 

\mypar{PRB budgeting}
A practical issue is that determining the PRB budget requires both the target SM count and the slot's average MCS. 
However, the true average MCS is known only after PRB allocation completes, creating a circular dependency: the scheduler must cap PRBs before allocation, yet the required lookup-table input is finalized only afterwards.

To break the MCS-dependency cycle, \sys{} bootstraps scheduling using the previous slot's average MCS.
At the start of each slot, and on each subsequent re-entry iteration, the scheduler pairs an MCS estimate with the target SM count and queries the lookup table to obtain a PRB budget. 
We define \emph{average MCS} as the PRB-weighted average of per-UE MCS values. 
The initial MCS estimate and target SM count are inherited from the previous slot's scheduling decision; on re-entry, they are updated using the true average MCS and adjusted SM count from the preceding iteration. 
Standard proportional-fair scheduling~\cite{openairinterface} then proceeds within the resulting PRB budget, preserving UE fairness while respecting the target SM reservation.

\subsubsection{Adaptive SM reservation adjustment}

The second issue is deciding when and by how much to adjust the SM reservation. 
Frequent adjustments expose slot-level traffic fluctuations directly to co-located workloads (FM training in our case), reducing the effective compute budget available to those workloads. 
Conversely, too coarse-grained adjustments may waste harvestable compute during low load or under-provision the RAN during sustained traffic increases. 
So, the scheduler must balance stability with responsiveness.

To achieve this, \sys{} continuously monitors RAN backlog and QoS indicators. 
Rather than reacting to every transient traffic burst, it absorbs short-lived fluctuations within UE buffers and adjusts the SM reservation only when backlog growth approaches a QoS-aware threshold. 
Given the uplink PRB scheduling decisions, the scheduler examines per-UE backlogs and the current SM reservation:
\begin{itemize}[leftmargin=*, nosep]
	\item \emph{Hold:}
	      If all UE backlogs are empty or below a tolerance threshold, the SM reservation stays unchanged.

	\item \emph{Ramp-up:}
 If any UE backlog exceeds the pre-defined threshold, the scheduler immediately increases the SM count to drain the UE buffer backlog within the current slot, preserving UE performance.

	\item \emph{Ramp-down:}
	      If no ramp-up occurs, the scheduler checks whether the current slot could have been served with fewer SMs.
	      A hysteresis counter tracks sustained over-provisioning and triggers a gradual step-down when appropriate.
	      A ``fast-exit'' path applies when the current slot allocated zero PRBs, allowing prompt down-sizing without waiting for hysteresis, so that spare SMs can be released to co-located workloads.
\end{itemize}

This policy is intentionally asymmetric: ramp-up is immediate to protect RAN performance, whereas ramp-down is conservative to improve stability. 
Together, the hold and hysteresis mechanisms suppress unnecessary SM reservation fluctuations, while fast-exit avoids prolonged over-provisioning and quickly updates the reservation to release spare SMs.

\subsubsection{Iteration control with PRB allowance check}

After PRB allocation and SM adjustment, both the average MCS and the target SM reservation may deviate from the values assumed for PRB budgeting.
To this end, the scheduler recomputes the PRB allowance (i.e., maximum supported PRB count) by querying the lookup table (\cref{fig:sm-lookup-table}c) with these two updated values.
If the allocated PRBs are within a close margin \emph{below} this allowance, the grants are directly emitted. 
Otherwise, the scheduler loops back to PRB budgeting with the updated MCS and SM count, repeating steps \textcircled{a}--\textcircled{c} in~\cref{fig:ran-control} until PRB allocation converges and the current-slot UL PRB grant is generated.
A small iteration bound (i.e., five) safeguards against non-convergence; if exceeded, the scheduler forces an SM ramp-up to guarantee forward progress.

By combining all three mechanisms above, our compute-aware scheduler determines the final UL PRB grant for each UE while deriving the current-slot SM reservation for the RAN and publishing the remaining SMs as spare.
Over time, a stable SM reservation is produced, allocating only the minimum necessary SM count per slot to guarantee RAN performance while maximizing spare-compute opportunity. Algorithm details are provided in Appendix~\ref{app:ran-ctrl-algo}, with parameter settings and their tuning in Appendix~\ref{app:ran-ctrl-params}.
    
\section{Two-Level Elastic Training}
\label{sec:training-strategy}

The controller in \cref{sec:ran-control} yields a stable SM reservation for the RAN and consequently makes spare-compute budget relatively stable within a cell site. But from a FM training perspective, available spare compute still evolves over time (see \cref{fig:sm-impact-on-training}b) and is also heterogeneous across cell sites. 
Static distributed training configurations are therefore inefficient: workers with insufficient spare compute become stragglers, while workers with insufficient work under-utilize available spare compute. 
Note that we consider distributed FM training using standard PP+DP training strategy as with modern LLM frameworks~\cite{dtfm,asteroid,confidant}: the model is partitioned into pipeline stages (PP), each replicated across workers via data parallelism (DP), so the runtime views a job as stage tasks connected by activation transfers and synchronized through data-parallel gradient exchange.

\begin{figure}[t]
	\centering
	\includegraphics[width=0.95\linewidth]{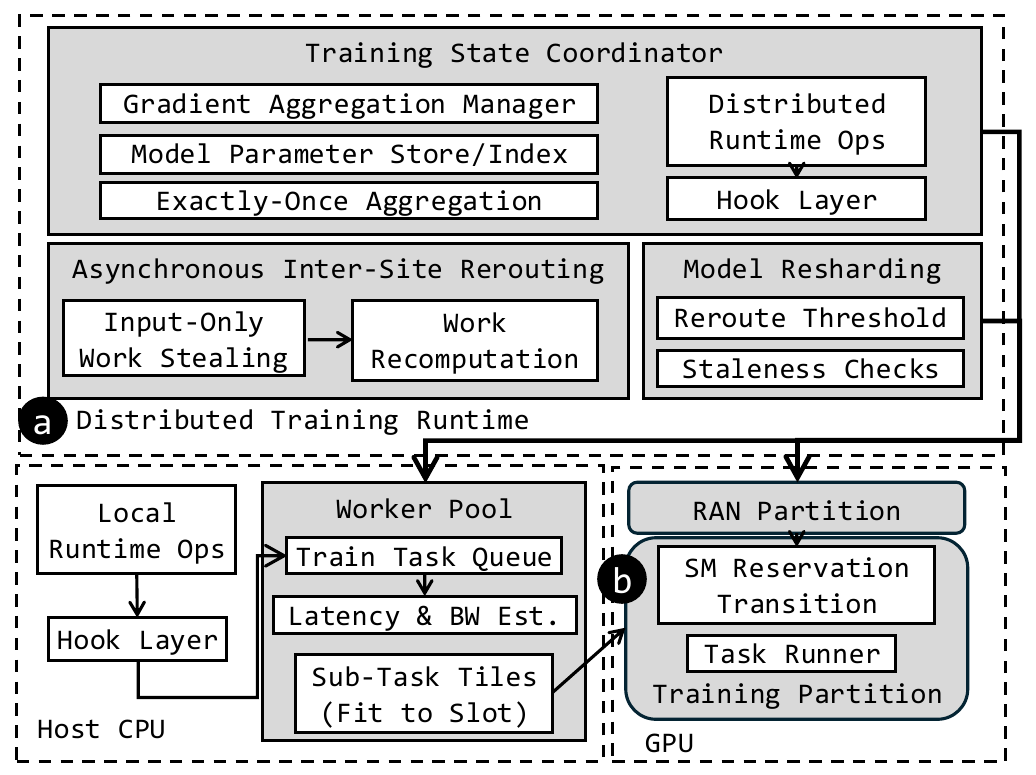}
	\caption{Two-level elastic training architecture: the inter-site coordinator manages AIR rerouting and MLR rebalancing across RAN sites (top), while each intra-site scheduler fits RAN-slot-bounded GEMM tiles into RAN-slot-level spare compute (bottom).}
	\label{fig:green_ctx_runtime}
\end{figure}

To cope with diverse and time varying spare compute across cell sites, \sys{} dynamically distributes the FM training computation across workers adapting with the spare-compute budget exposed by the RAN. Specifically, \sys{} employs a two-level elastic training framework (\cref{fig:green_ctx_runtime}) that adapts to spare-compute availability both within a site and across sites:
(i)~\emph{inter-site scheduling} (\textcircled{a}, \cref{sec:two-level-scheduling}): harvests capacity across sites at slower timescales, from sub-second \gls{air} decisions to hourly traffic shifts via \gls{mlr}; and
(ii)~\emph{intra-site scheduling} ( \textcircled{b}, \cref{sec:absorbing-local-fluctuation}): adapts to RAN-slot-level variation locally, and exposes an aggregated capacity view to the inter-site coordinator.
This two-level approach is essential because RAN-slot-level fluctuations ($<$1\,ms) are too fast for inter-site coordination ($>$100\,ms), while macro-scale shifts are too slow-varying\footnote{Note that an hour spans millions of RAN slots.} to be handled efficiently by intra-site scheduling alone. Note that \gls{mlr} is also termed as model resharding as in \autoref{fig:green_ctx_runtime}.

\subsection{Inter-Site Scheduling}
\label{sec:two-level-scheduling}

Inter-site scheduling must cope with model and optimizer states on the order of 10--100\,GB~\cite{deepspeed} at each cell site over 1--10\,Gbps inter-site links~\cite{gsma-backhaul,gsma-5g-cost}, so migrating states would take seconds to minutes.
Because RAN cell sites connect to the mobile network operator's core infrastructure in a star topology with site-to-core links that have 10--50$\times$ higher bandwidth inter-site links~\cite{gsma-backhaul,nokia-anyhaul}, we use a coordinator-centric runtime situated at the operator's core network infrastructure for lightweight control-plane coordination while minimizing data-plane state transfers between sites.

\mypar{Centralized training state coordinator}
The inter-site coordinator tracks whereabouts of model state and in-flight training work, and maintains per-sample-stage commit records. 
This is complemented by \gls{air} that reroutes micro-batch stage tasks, and \gls{mlr} that reshapes the model-parallel layout.
The coordinator couples to the training runtime through lightweight hooks at existing synchronization points and derives latency signals from task arrivals and completions at site boundaries.
More details in Appendix~\ref{app:scheduling-mechanism-details}.

\mypar{Asynchronous inter-site rerouting (AIR)}
\gls{air} seeks to balance load across sites at sub-second timescales by rerouting micro-batch stage tasks (i.e., a set of GEMM tiles) while keeping model state pinned at each site.
The underlying principle is \emph{move computation work, not model state}.
At each AIR control interval, the coordinator assigns queued micro-batch stage tasks to equivalent-stage executors (workers) whose predicted completion time fits within the current training-step admission deadline.
If an executor can no longer meet that deadline, \gls{air} reroutes a pending micro-batch stage task to another equivalent-stage executor that holds the same model parameters; when no replica can finish before the training step closes, the not-yet-owned task is deferred for admission in the next training step.
Only inputs or hidden states move between sites, and gradients are recomputed when necessary.

\begin{figure}[t]
	\centering
	\includegraphics[width=\linewidth]{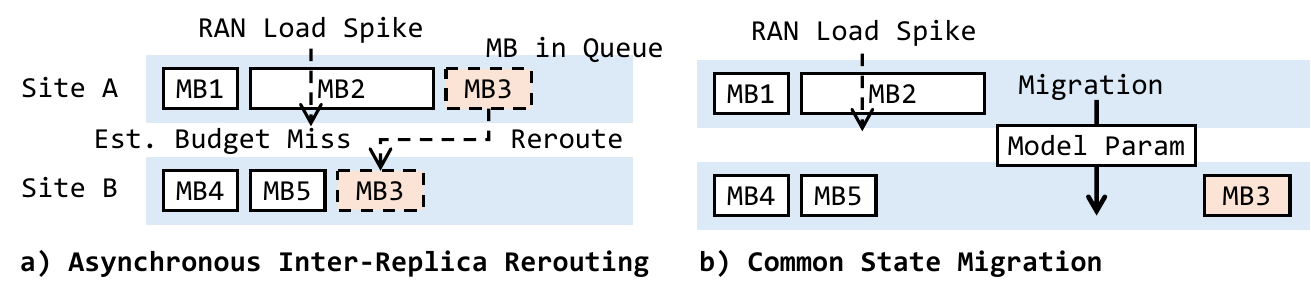}
	\caption{An example illustrating the benefit of AIR compared against common state migration (MB denotes Micro-Batch).}
	\label{fig:air}
\end{figure}

Consider a two-site example in \cref{fig:air} with two equivalent-stage executors in different data-parallel replicas.
If a spike in RAN load slows site~A while it executes MB2, \gls{air} stops assigning new micro-batch stage tasks to site~A and reroutes the next queued micro-batch stage task (MB3 in the example) to site~B, while letting MB2 finish on site~A.
If MB2 is the last micro-batch of the current training step, \gls{air} speculatively re-executes its stage task on site~B and accepts only the first valid result for each sample-stage key.
Distinct from the migration-based approaches (illustrated in \cref{fig:air}b), \gls{air} reacts without waiting for ongoing work to drain and moves only KB--MB-scale inputs or activations, not multi-GB model state.

\mypar{Model layout rebalancing (MLR)}
\gls{air} absorbs transient fluctuations within a fixed model-parallel layout. 
But when one site remains persistently slower, deferred micro-batches accumulate and rerouting alone cannot remove the structural imbalance.
The coordinator therefore tracks the per-training-step deferral fraction with a moving average and escalates to \gls{mlr} only when it stays above a threshold across consecutive training steps (illustrated in Appendix \cref{fig:cross-site-deferral-trigger}), so this mechanism is applied only for persistent spikes and imbalances.
When \gls{mlr} fires, the pipeline-parallel partition boundary shifts at a training-step barrier so that overloaded sites shed layers to less-loaded sites; only boundary-crossing layers' parameters and optimizer state move, via a \textsc{Prepare/Copy/Commit} sequence whose drain phase alone stalls training (measured breakdown in Appendix Table~\ref{tab:reshard-breakdown-ts}).
The full rebalancing workflow and fault-tolerance support appear in Appendix~\ref{app:scheduling-mechanism-details}.

\mypar{Correctness}
\gls{air} and \gls{mlr} affect only the execution location and timing of each workload. They do not change which samples contribute to training or how their gradients are aggregated.
The training correctness guarantees (including at-most-once fenced commits, weighted aggregation matching a synchronous run, bounded debt) are stated and proved in Appendices~\ref{app:correctness}--\ref{app:air-proofs}.

\subsection{Intra-Site Scheduling}
\label{sec:absorbing-local-fluctuation}

Next, we explain how \sys{} adapts to available spare compute (SMs) within a single cell site.
Here the challenge arises because FM training kernels are inherently long-running, whereas spare compute is exposed at RAN-slot granularity. 
Since FM training is dominated by GEMM operations (\S\ref{sec:background}), \sys{} achieves intra-site elasticity by controlling their execution granularity. 
However, individual GEMMs as a whole can be computationally heavy and may need multiple RAN slots, making them poorly matched to the rapidly changing spare SMs exposed by the spare-compute controller. 
A GEMM that fits within the available SM budget in a slot may no longer fit if the RAN expands its SM reservation in the next slot. 
To address this, \sys{} decomposes GEMMs into fine-grained \emph{GEMM tiles}, which become the basic schedulable unit of computation and can be paused, resumed, and migrated with minimal disruption.

\mypar{RAN-slot-bounded GEMM tiles}
As per the above, the intra-site scheduler queues the GEMM tiles assigned by the inter-site scheduler and admits them against each RAN slot's spare SM and bandwidth budgets, so training compute stays bounded and safely co-exists with the RAN workload. 
A smoothed window of aggregate spare capacity is reported back to the inter-site scheduler (\cref{sec:two-level-scheduling}).
As outlined above, \sys decomposes \glspl{gemm} into GEMM tiles, and admits them only when they satisfy two constraints under the current RAN-slot spare-SM budget:
(i) predicted completion before the end of the current RAN slot, and
(ii) required memory fits within the remaining memory/interconnect headroom.
\sys further uses persistent kernels~\cite{gupta2012persistent,osama2023streamk}.
     
\section{Evaluation}
\label{sec:evaluation}

We evaluate \sys{} along three dimensions. 
First, we determine whether its RAN-centric spare-compute controller stabilizes the RAN SM reservation while preserving RAN performance, and whether the resulting stable spare-compute budget improves co-located training throughput (\cref{sec:eval-single-site}). 
Second, we evaluate whether \sys{}'s two-level elastic training framework efficiently harvests heterogeneous and time-varying spare compute across sites, and study its scalability to larger deployments and models (\cref{sec:eval-real-training}). 
Finally, we quantify \sys{}'s runtime overheads (\cref{sec:eval-overheads}).

\subsection{Effectiveness of RAN-Centric Spare Compute Control}
\label{sec:eval-single-site}

We first evaluate \sys{}’s RAN-centric spare compute controller on a live 5G NR stack, focusing on RAN performance, compute stability, and benefits to co-located training.

\mypar{\textit{Testbed}}
Unless otherwise stated, we use the OAI~\cite{openairinterface} 5G stack with GPU-accelerated LDPC decoding~\cite{openairinterface-gpu} on a 48-SM DGX Spark, using band n78 (3.5~GHz TDD, 40~MHz), for the experiments conducted in this section.

\mypar{\textit{Baselines}}
\textsc{NoCtrl} uses the unmodified OAI MAC scheduler and exposes its raw SM
demand without shaping. 
Both \textsc{NoCtrl} and \sys{} use CUDA Green Contexts to enable co-location with the training workload.
\textsc{YinYangRAN}~\cite{yinYangRAN} represents MPS-based GPU sharing:
we grant it oracle knowledge of future RAN demand and size each one-second partition for the peak requirement within that interval.
We emulate MPS for the RAN workload rather than enforcing MPS directly, since MPS reconfiguration requires CPU fallback and disrupts RAN operation on our testbed.

\subsubsection{Impact on RAN Performance and SM Reservation}
\label{sec:eval-ran}

\begin{table}[t]
	\centering
	\small
	\setlength{\tabcolsep}{3pt}
	\caption{OTA results (2~UEs): iperf (5\,min bidirectional); WebRTC call quality (10\,min bidirectional).}
	\label{tab:ota-validation}
	\clampbox{\linewidth}{
		\begin{tabular}{llccc}
			\toprule
			\textbf{Test} & \textbf{Metric}      & \textbf{\sys} & \textbf{NoCtrl} & \textbf{$\Delta$\,(\%)} \\
			\midrule
			iperf         & Throughput (Mbps)    & 12.1          & 12.2            & $-$0.8                  \\
			\midrule
			WebRTC        & Avg.\ bitrate (kbps) & 2440          & 2460            & $-$0.8                  \\
			              & Packet loss (\%)     & 0.35          & 0.33            & $+$6.1                  \\
			              & Jitter (ms)          & 3.4           & 3.3             & $+$3.0                  \\
			              & RTT (ms)             & 22.4          & 22.8            & $-$1.8                  \\
			              & Freeze count         & 0             & 1               & ---                     \\
			\bottomrule
		\end{tabular}
	}
\end{table}

\begin{table}[t]
	\centering
	\small
	\caption{Aggregated application-layer KPIs (3~UEs, 90\,s UL UDP via D-ITG w/ real-world channel replay).}
	\label{tab:ue-app-kpi}
	\clampbox{\linewidth}{
		\begin{tabular}{lccc}
			\toprule
			\textbf{Metric}        & \textbf{\sys} & \textbf{NoCtrl} & \textbf{$\Delta$\,(\%)} \\
			\midrule
			Throughput, sum (Mbps) & 4.56          & 4.56            & $+$0.1                  \\
			Delay, mean (ms)       & 18.4          & 19.0            & $-$3.1                  \\
			Delay, p95 (ms)        & 120.2         & 124.3           & $-$3.3                  \\
			Jitter, mean (ms)      & 4.4           & 4.5             & $-$3.0                  \\
			Jitter, p95 (ms)       & 35.9          & 38.1            & $-$5.9                  \\
			Packet loss (\%)       & 66.7          & 66.7            & $-$0.03                 \\
			\bottomrule
		\end{tabular}
	}
\end{table}

\mypar{{Impact on application performance}}
First, we use \texttt{iperf3} to measure sustained throughput and WebRTC~\cite{webrtc} calls to evaluate interactive traffic over the air (OTA) using a USRP B210 and two 5G phones. 
Then, we use D-ITG~\cite{ditg} to generate bursty UDP traffic for three OAI \texttt{RFSIM} UEs to measure fine-grained network KPIs while varying the channel conditions based on the production O-RAN traffic~\cite{tractor}. More evaluation details are provided in Appendix~\ref{app:ran-eval-methodology}.

The results show that \sys{} preserves application performance. 
In OTA experiments (\cref{tab:ota-validation}), the \texttt{iperf3} throughput differs by less than 1\% from \textsc{NoCtrl} while WebRTC performance is also comparable.
As for the D-ITG workload (\cref{tab:ue-app-kpi}), \sys{} and \textsc{NoCtrl} achieve the same aggregate throughput and packet-loss rate, while mean and p95 delay and jitter remain comparable. 
Overall, these results show that \sys{}'s compute-aware scheduling does not
degrade application performance.

\begin{table}[t]
	\centering
	\caption{
    SM reservation stability and HARQ deadline-met rate in two TRACTOR scenarios (4 UEs, 600\,s). 
    $\uparrow$/$\downarrow$ means the higher/lower the better.
    }
	\label{tab:ran-ablation-overhead}
	\clampbox{0.9\linewidth}{
		\setlength{\tabcolsep}{3pt}
		\begin{tabular}{llcc@{\,}cc}
			\toprule
			\textbf{Scen.}            & \textbf{Method}     & \textbf{Mean SM cnt.$\downarrow$} & \multicolumn{2}{c}{\textbf{SM trans.$\downarrow$}} & \textbf{ddl.\ met$\uparrow$}          \\
			\midrule
			\multirow{3}{*}{Scen.\,1} & \textsc{NoCtrl}     & 14.2                  & 435                                    & (0.7/s)            & 99.8\% \\
			                          & \textsc{YinYangRAN} & 34.1                  & 96                                     & (0.16/s)           & --     \\
			                          & \sys                & 14.4                  & 89                                     & (0.1/s)            & 99.8\% \\
			\midrule
			\multirow{3}{*}{Scen.\,2} & \textsc{NoCtrl}     & 8.4                   & 744                                    & (1.2/s)            & 99.8\% \\
			                          & \textsc{YinYangRAN} & 22.5                  & 124                                    & (0.21/s)           & --     \\
			                          & \sys                & 8.5                   & 187                                    & (0.3/s)            & 99.8\% \\
			\bottomrule
		\end{tabular}
	}
\end{table}

\begin{figure}[t]
	\centering
	\includegraphics[width=0.95\linewidth]{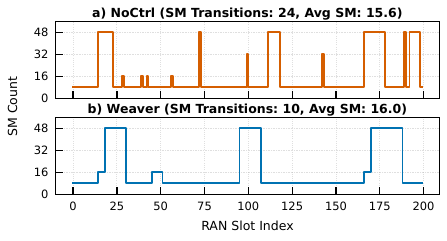}
	\caption{SM reservation over time. 
    For brevity, we only show the case of a sample period from Scenario 2.}
	\label{fig:sm-timeline}
\end{figure}

\mypar{{Impact on SM reservation stability}}
Next, we evaluate whether \sys{} can effectively stabilize the RAN SM reservation and thereby expose a more stable spare-compute budget.
We replay two four-UE TRACTOR 5G scenarios (production O-RAN traffic)~\cite{tractor} in OAI \texttt{RFSIM} for 600\,s, while varying the channel conditions accordingly:
static UEs with steady traffic (Scen.~1, Trial2) and walking UEs with dynamic traffic (Scen.~2, Trial3). 

\cref{tab:ran-ablation-overhead} reports the mean RAN SM reservation, number
of SM transitions, and HARQ deadline-met rate.
Compared with \textsc{NoCtrl}, \sys{} reduces SM transitions by
$4.0$--$4.9\times$, while maintaining nearly the same mean SM reservation
and the same 99.8\% deadline-met rate.
\textsc{YinYangRAN} achieves comparable transition counts but reserves
$2.4$--$2.7\times$ more SMs because each partition must accommodate the peak
demand within its interval.
Thus, consistent with our design in \cref{sec:ran-control}, \sys{} effectively
smooths short-term RAN demand into longer stable SM-reservation intervals, as further substantiated by the timeline in \cref{fig:sm-timeline}.

\subsubsection{Impact on Usable Spare Compute}
\label{subsec:e2e-prototype}

We next evaluate whether the more stable SM reservations produced by \sys{} translate into more usable compute for a co-located workload.

\begin{table}[t]
\centering\small
\caption{Useful throughput of 64-GEMM tasks during RAN coexistence (four UEs, 120\,s per case).}
\label{tab:e2e-prototype}
\clampbox{\linewidth}{
\setlength{\tabcolsep}{6pt}
\begin{tabular}{lrr}\toprule
& \multicolumn{2}{c}{Useful throughput (TFLOPS)} \\
\cmidrule(lr){2-3}
Method & Scenario 1 & Scenario 2 \\\midrule
\textsc{YinYangRAN} & 0.183 & 0.183 \\
\textsc{NoCtrl} & 0.481 & 0.779 \\
\sys{} & \textbf{1.398} & \textbf{1.374} \\
\bottomrule\end{tabular}}
\end{table}

\mypar{{Controlled compute workload}}
Using the same RAN setup and TRACTOR traces (\cref{sec:eval-ran}), \emph{we run the live end-to-end RAN workload, including actual UE transmissions through the OAI stack, while co-locating a synthetic training workload on the same GPU}.
The workload consists of repeated chains of 64 dependent GEMMs; let $M$ and $N$ denote the output matrix dimensions, while $K$ is the shared reduction dimension in $C_{M\times N}=A_{M\times K}B_{K\times N}$, and we have $M=20480, N=K=1024$. This allows us to evaluate the impact of RAN control on useful computation while keeping the training-side execution identical across methods.

As shown in \cref{tab:e2e-prototype}, \sys{} sustains approximately 1.4\,TFLOPS in
both scenarios, achieving $2.91\times$ and $1.76\times$ the useful
throughput of \textsc{NoCtrl}, and $7.50$--$7.62\times$ that of
\textsc{YinYangRAN}.
This gain follows directly from the more stable reservations while not degrading RAN performance, as observed in
\cref{sec:eval-ran}: fewer SM transitions provide longer uninterrupted intervals in which the co-located workload can make useful progress.

\begin{table}[t]
	\centering
	\caption{L40S training under recorded RAN SM reservations. $\uparrow$/$\downarrow$ means the higher/lower the better.}
	\label{tab:qualitative-comp}
	\clampbox{\linewidth}{
		\setlength{\tabcolsep}{3pt}
		\begin{tabular}{rcccc}
			\toprule
			\textbf{Method}     & \textbf{Violations} $\downarrow$ & \textbf{Efficiency} $\uparrow$ & \textbf{Token/s} $\uparrow$ & \textbf{TFLOPS} $\uparrow$ \\
			\midrule
			\textsc{YinYangRAN} & N/A                              & 32\%                           & 1447                        & 54                         \\
			\textsc{NoCtrl}     & 2.7\%                            & 67\%                           & 3827                        & 118                        \\
			\sys                & \textbf{0.6\%}                   & \textbf{83\%}                  & \textbf{4088}               & \textbf{142}               \\
			\bottomrule
		\end{tabular}
	}
\end{table}

\mypar{{End-to-end FM training workload}}
We next verify that this benefit carries over to actual FM training.
Here, we use a data-center grade GPU, an NVIDIA L40S, and replay the earlier RAN SM-reservation traces (rescaled to the L40S SM count) to train a OPT-13B~\cite{opt} configured with 12 layers (with target global batch size $B^\star=4$ and sequence length 512) under the same reservation traces.
All settings use an identical model and optimizer state.
For \textsc{YinYangRAN}, each reconfiguration checkpoints training state
to host memory.

As shown in~\cref{tab:qualitative-comp}, \sys{} harvests 83\% of the available spare compute, compared with 67\% for \textsc{NoCtrl} and 32\% for \textsc{YinYangRAN}.
This translates to 4088 tokens/s and 142\,TFLOPS, corresponding to $1.07\times$ higher token throughput and $1.20\times$ higher compute throughput than \textsc{NoCtrl}.

Interestingly, the higher training throughput also comes with better allocation compliance.
An isolation violation occurs when a training kernel still occupies SMs required by the RAN at decoding start.
\sys{} reduces the violation rate from 2.7\% under \textsc{NoCtrl} to 0.6\%, corresponding to 118 versus 26 events, while reducing their aggregate duration by $2.7\times$.
This shows that the more stable RAN reservation not only exposes more usable compute, but also allows the training workload to adapt more cleanly to the RAN's changing compute demand.

\subsection{Effectiveness of Two-Level Elastic Training}
\label{sec:eval-real-training}

Having shown that \sys{} exposes more usable spare compute at RAN sites, we next evaluate whether its two-level elastic training framework can efficiently harvest this heterogeneous and time-varying compute across sites.
We first evaluate end-to-end training on our eight-site GPU testbed, then isolate its adaptation to dynamic spare compute, and finally study scalability to larger deployments and models.

\mypar{\textit{Testbed}}
Unless otherwise stated, we evaluate \sys{} on an eight-site GPU-backed 5G testbed, with one NVIDIA L40S GPU (48\,GB) at each site.
The sites are distributed across different cities; measured inter-site bandwidth and latency are reported in Appendix~\ref{app:eval-settings}, and implementation details are provided in Appendix~\ref{app:implementation}.
We replay the SM-reservation traces from \cref{sec:eval-ran}, rescaled to the L40S SM count, to reproduce millisecond-scale RAN compute dynamics.

\mypar{\textit{Models and datasets}}
We use the OPT family throughout the training evaluation.
Our physical-testbed experiments use OPT-125M–6.7B, while the large-scale simulation in \cref{sec:eval-sim} evaluates OPT-13B–70B.
We train OPT on SST-2~\cite{sst2} with global batch size $B^\star=32$ and sequence length 512.
The training scheduler determines the micro-batch size based on the data-parallel (DP) and pipeline-parallel (PP) configuration.

\mypar{\textit{Baselines}}
We compare against three distributed training systems designed for heterogeneous and time-varying resources:
(i)~\textsc{DTFM}~\cite{dtfm}, which uses DP+PP with coarse-grained checkpoint-and-restore;
(ii)~\textsc{Asteroid}~\cite{asteroid}, which uses hybrid PP and pipeline replay to tolerate stragglers; and
(iii)~\textsc{Confidant}~\cite{confidant}, which combines local updates with periodic global aggregation to tolerate slow or unavailable sites.
In our evaluation, these baselines use MPS for GPU sharing, while \sys{} uses CUDA Green Contexts.

\subsubsection{End-to-End Training Performance}

We first evaluate the end-to-end benefit of \sys{}'s two-level elastic training framework under heterogeneous and time-varying spare compute across multiple RAN sites.

We evaluate the full \sys{} stack against each baseline under the same SM-reservation trace replay for 1000 training steps, covering both per-site capacity fluctuations and longer-term inter-site drift.
We report token throughput as the end-to-end training metric and achieved compute throughput, normalized by aggregate peak cluster compute throughput (in terms of TFLOPS), as a measure of how effectively available compute is converted into useful training work.

\begin{figure}[t]
	\centering
	\includegraphics[width=\linewidth]{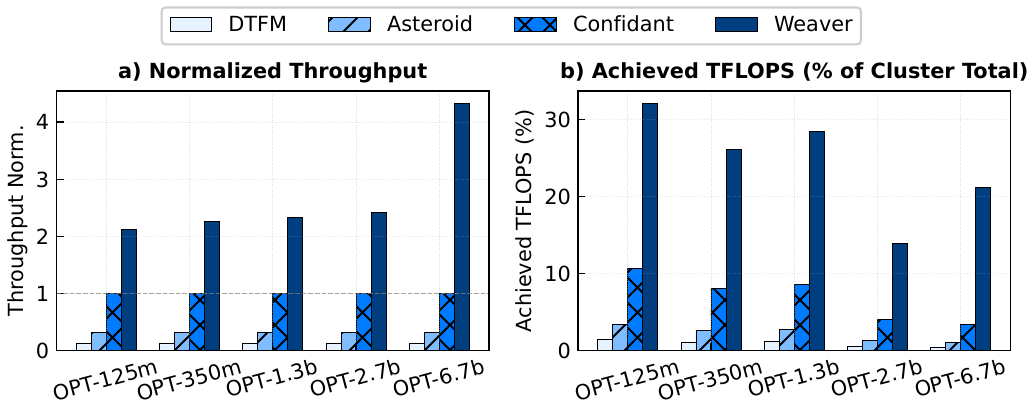}
	\caption{Training throughput and achieved TFLOPS as a percentage of peak cluster compute across OPT model sizes.}
	\label{fig:model-scaling}
\end{figure}

\mypar{{Training throughput}}
From the results in~\cref{fig:model-scaling}a, we observe that \sys{} consistently achieves higher end-to-end training throughput across the evaluated model sizes.
For OPT-1.3B, \sys{} achieves 35K tokens/s, $2.1\times$ that of the strongest baseline, \textsc{Confidant} (17K tokens/s). 
The gap widens to $3.7\times$ for OPT-6.7B (6.3K versus 1.7K tokens/s).
\sys{} also consistently achieves a higher fraction of peak cluster compute throughput across model sizes (\cref{fig:model-scaling}b).
Together, the higher token throughput and compute utilization show that \sys{} more effectively converts dynamic spare compute into useful training work.
In contrast, the baselines rely on coarser adaptation mechanisms that are less effective under dynamic and heterogeneous resource availability.

\mypar{{Training stability and convergence}}
Beyond mean throughput, \sys{} also maintains stable training under replayed
RAN dynamics, with lower step-level throughput variation and fewer throughput
drops than the baselines.
In our experiments, the training-loss trajectory closely tracks a fixed-cluster synchronous SGD reference, indicating that \sys{}'s dynamic adaptation does not compromise training convergence.
We provide more detailed stability and convergence results in Appendix~\ref{app:stability-convergence}.

\subsubsection{Adaptation to Dynamic Spare Compute}

We next isolate how effectively \sys{}'s inter-site scheduler adapts training to dynamic spare compute across sites.
To remove differences from intra-site execution, we enable the same intra-site scheduler for all methods and initialize them with the same model-parallel partition.
Thus, the remaining performance difference comes from how each method adapts training across sites as spare compute changes.

Over a 6-minute trace with 41 capacity-drop events across eight sites, \sys{} sustains 15--32\% of peak cluster TFLOPS, compared with 6--11\% for \textsc{Confidant} and 1--2\% for \textsc{DTFM}.
This shows that \sys{} effectively preserves useful training computation as capacity shifts across sites.

\mypar{{Inter-site adaptation behavior}}
We zoom into the scheduler internals to understand how it reacts to these dynamics.
Over the same 6-minute replay, spare compute shifts 80 times at the training-step timescale.
The \gls{air} fast-path handles 86.2\% of these fluctuations, by rerouting pending micro-batch stage tasks, without invoking model layout rebalancing (MLR), with a median exposed stall of 66\,ms (p95: 68\,ms).
Only 11 events escalate to the \gls{mlr} slow-path.
For these events, the only exposed critical-path cost is the 528\,ms PREPARE drain; model-state transfer and layout activation overlap with useful
training on unaffected stages.
Together, these results validate the two-timescale design in
\cref{sec:two-level-scheduling}: \gls{air} absorbs most transient capacity changes, while
\gls{mlr} is invoked only for persistent imbalance.
Additional mechanism-level results and parameter sensitivity are reported in
Appendix~\ref{sec:eval-inter-scheduling}.

\subsubsection{Scalability to Larger Deployments and Models}
\label{sec:eval-sim}

Finally, we evaluate whether \sys{}'s training design continues to scale beyond the size of our physical testbed, both to larger RAN deployments and larger foundation models.

\mypar{\textit{Environment settings}}
We evaluate 128--512-site deployments, all with 48\,GB GPUs, under two RAN settings.
\emph{Rescaled} scales our measured testbed topology while preserving its link characteristics.
\emph{Typical} uses a representative RAN star topology with 32 sites per core, 25\,Gbps/1\,ms site-to-core, 400\,Gbps/20\,ms core-to-core, and 10\,Gbps peer links~\cite{gsma-backhaul,nokia-anyhaul}.

\mypar{\textit{Methodology}}
We use a profile-based simulator built on Morphling~\cite{morphling}.
Its inputs combine measured per-layer compute profiles from truncated-depth training runs, inter-site bandwidth and latency, per-link contention, per-site memory constraints, and TRACTOR-derived SM-reservation traces; further details are provided in Appendix~\ref{app:sim-methodology}.
The simulator does not model site failures or stragglers beyond their effect on the available SM reservation.
We compare against \textsc{Confidant}, the strongest baseline in our physical testbed, and \textsc{Confidant*}, which strengthens it with exhaustive DP+PP configuration search and perfect spare-compute harvesting.
This separates \sys{}'s scaling benefits from limitations due to baseline configuration or compute harvesting.

\begin{table}[t]
    \centering\small
    \caption{{OPT-13B training throughput (K tokens/s) as site count increases; 48\,GB GPUs.}}
    \label{tab:site-scaling-sim}
    \clampbox{\linewidth}{{\begin{tabular}{l rrr rrr}
            \toprule
            & \multicolumn{3}{c}{\emph{Rescaled}} & \multicolumn{3}{c}{\emph{Typical}} \\
            \cmidrule(lr){2-4} \cmidrule(lr){5-7}
            \textbf{System} & \textbf{128} & \textbf{256} & \textbf{512} & \textbf{128} & \textbf{256} & \textbf{512} \\
            \midrule
            \sys{}     & \textbf{76.8} & \textbf{87.8} & \textbf{92.2} & \textbf{106.7} & \textbf{198.8} & \textbf{350.6} \\
            Confidant* & 47.6 & 47.6 & 47.6 & 58.5 & 77.6 & 87.3 \\
            Confidant  & 46.6 & 46.6 & 46.6 & 46.5 & 48.6 & 49.7 \\
            \bottomrule
        \end{tabular}}}
\end{table}

\mypar{{Scaling across sites}}
\cref{tab:site-scaling-sim} shows that \sys{} continues to benefit from additional sites despite heterogeneous spare compute and constrained inter-site connectivity.
Under \emph{Typical}, OPT-13B throughput increases from 106.7K to 350.6K tokens/s as the deployment grows from 128 to 512 sites, a $3.29\times$ increase, compared with $1.49\times$ for \textsc{Confidant*}.
At 512 sites, \sys{} achieves $4.0\times$ the throughput of \textsc{Confidant*} and $7.1\times$ that of \textsc{Confidant}.
Scaling is more constrained under the \emph{Rescaled} interconnect: \sys{} increases from 76.8K to 92.2K tokens/s, while both baselines remain nearly flat.
These results show that \sys{} can effectively exploit large-scale distributed spare compute.

\mypar{{Scaling to larger models}}
We next evaluate whether \sys{} can extend training to models that exceed the sizes feasible on our physical testbed.
On 512 sites, \sys{} fits OPT-30B and OPT-70B within the 48\,GB per-site memory budget, with peak memory footprints of 30.2\,GB and 40.2\,GB, respectively (\cref{tab:model-scaling-sim}).
In contrast, both \textsc{Confidant} and \textsc{Confidant*} exceed the per-site memory budget for both models.
Under \emph{Typical}, \sys{} sustains 146.1K tokens/s for OPT-30B and 58.4K tokens/s for OPT-70B.
Thus, \sys{} remains capable of harvesting distributed RAN spare compute as both deployment size and model size increase.

\begin{table}[t]
    \centering
    \caption{{Larger-model feasibility and throughput on 512 sites with 48\,GB GPUs. OOM~= out of memory.}}
    \label{tab:model-scaling-sim}
    \clampbox{\linewidth}{{\begin{tabular}{llrrr}
            \toprule
            & & \textbf{Peak mem.} & \multicolumn{2}{c}{\textbf{K tokens/s}} \\
            \cmidrule(lr){4-5}
            \textbf{Model} & \textbf{System} & \textbf{(GB)} & \textbf{Rescaled} & \textbf{Typical} \\
            \midrule
            OPT-30B & \sys{} & 30.2 & \textbf{38.4} & \textbf{146.1} \\
                    & Confidant* & 57.6 & OOM & OOM \\
                    & Confidant & 70.0 & OOM & OOM \\
            \midrule
            OPT-70B & \sys{} & 40.2 & \textbf{15.4} & \textbf{58.4} \\
                    & Confidant* & 75.2 & OOM & OOM \\
                    & Confidant & 149.8 & OOM & OOM \\
            \bottomrule
        \end{tabular}}}
\end{table}

\subsection{Runtime Overheads}
\label{sec:eval-overheads}

We quantify the runtime overheads introduced by \sys{} across both the RAN and training control paths.
\sys{} introduces modest runtime overheads.
The introduction of the RAN-centric spare compute controller into the MAC scheduler incurs an 
overhead of 0.8\% ($0.13\mu\text{s}$) and 13.4\% ($6.56\mu\text{s}$) at the p50 and p99, respectively.
\sys{}'s training control plane requires only ${\sim}$120\,$\mu$s of CPU scheduling per training step, accounting for $<$0.3\% of step time, confirming that \sys control plane is lightweight.
To support rapid adaptation to SM-reservation changes, each worker pre-creates up to 16 Green Context streams, consuming 320\,MB of GPU memory ($<$0.7\% of a 48\,GB L40S), while rebinding the next GEMM tile to a pre-created Green Context takes less than 10\,$\mu$s (Appendix~\ref{app:implementation}).
Among sites, \gls{air} reroutes only KB--MB-scale inputs or activations rather than model state. 
These overheads are substantially smaller than the multi-GB state transfers and multi-second recovery costs incurred by coarse-grained inter-site scheduling.
Further implementation details are provided in Appendix~\ref{app:implementation}.
Broader deployment considerations, including full-L1 acceleration, applicability to other non-RAN workloads, tenant isolation, and energy/thermal constraints, are discussed in Appendix~\ref{app:practical-concerns}.

\section{Related Work}
\label{sec:related-work}

\mypar{Compute sharing for RAN}
We discuss recent GPU-sharing techniques for AI-and-RAN in~\cref{sec:background}; here, we focus on complementary work on sharing CPU resources in vRAN. 
Concordia~\cite{concordia} dynamically reallocates CPU cores across RAN and co-located workloads, while other systems use CPU quotas~\cite{OppCPUsharing} or fine-grained redirection and scheduling~\cite{Aquifer} to reclaim otherwise idle CPU capacity. 
These approaches benefit from fine-grained CPU thread preemption and migration, mechanisms that do not directly transfer to GPU execution. 
\sys{} instead targets GPU-accelerated RAN infrastructure, where resource sharing must be coordinated through SM allocation while preserving per-slot RAN deadlines.

\mypar{Characterizing RAN GPU Utilization}
Prior work has profiled the computational behavior of GPU-accelerated RAN processing. CloudRIC~\cite{CloudRIC} studies GPU utilization under different RAN loads, ETHOS~\cite{ethos} characterizes the performance and efficiency of virtualized O-RAN processing, and DecodeX~\cite{decodeX} benchmarks LDPC decoding across different compute platforms. 
These studies primarily focus on aggregate utilization (which is coarse-grained) or individual RAN components.
In contrast, our characterization is aimed specifically at understanding the spare GPU capacity available for co-located workloads. 
We therefore study GPU usage along multiple resource dimensions, i.e., SM occupancy, arithmetic utilization, and memory-bandwidth utilization, at RAN-slot granularity, and further examine how spare compute varies temporally and spatially across cell sites.

\mypar{Resource harvesting}
Volunteer-computing systems~\cite{boinc,folding-at-home} and harvested cloud VMs~\cite{harvest-vm} aggregate opportunistic idle resources for embarrassingly parallel workloads, typically treating capacity availability coarsely and scavenging slack over much longer timescales. 
In contrast, spare compute in GPU-accelerated RANs varies across fine-grained temporal, spatial, and resource dimensions, while the primary RAN workload must retain strict performance guarantees. 
These systems therefore do not address the fine-grained resource adaptation or cross-site coordination required for tightly coupled FM training.

\mypar{Dynamic ML training systems}
Existing training systems~\cite{dtfm,parcae,hap,confidant} adapt to heterogeneous or time-varying resources, but are not designed for the RAN setting, where GPU capacity can change at sub-millisecond timescales due to a high-priority, latency-critical co-tenant.
Other systems tolerate resource churn through comparatively heavyweight mechanisms, such as checkpoint-and-restart (\eg Bamboo~\cite{bamboo}, Mario~\cite{mario}) or recomputation (\eg SWARM~\cite{swarm}, Asteroid~\cite{asteroid}), which are poorly matched to such fine-grained resource dynamics.
Moreover, approaches such as per-device AllReduce in DTFM~\cite{dtfm}, and communication that grows with participating device count in Bamboo~\cite{bamboo} and Mario~\cite{mario}, can become costly over constrained inter-site links.
In contrast, \sys{} separates adaptation across timescales: fine-grained intra-site scheduling reacts to RAN-slot-level changes, while lightweight work rerouting and infrequent model-layout rebalancing handle cross-site dynamics.

\section{Conclusions}
\label{sec:conclusion}

This paper presents \sys, a system for safely and efficiently harvesting spare GPU compute in AI-RAN infrastructure for FM training.
\sys combines a RAN-centric spare-compute controller that stabilizes SM reservations with a two-level elastic training framework that adapts to resource variability across sites and timescales.
Our evaluation shows that \sys improves FM training throughput while preserving RAN performance.

\section*{Acknowledgements}
This work was supported in part by the UKRI/EPSRC grants UKRI860 and UKRI554, and the AI-RAN Alliance Innovation Award.

\balance
\bibliographystyle{abbrv}
\bibliography{references}

\clearpage
\appendix
\renewcommand{\sectionautorefname}{Appendix}
\section{Spare Compute Characterization}
\subsection{The LDPC Decoding Is Memory-Bound}
\label{app:spare-compute-character}
\label{app:ldpc-intensity}

We apply a roofline analysis to SionnaRK's~\cite{openairinterface-gpu} GPU-accelerated LDPC decoding implementation to show that the PHY kernels are decisively memory-bandwidth-bound. The relevant comparison point is the roofline ridge point,
\[
	\eta^* = 1000 \cdot P\,[\text{TFLOPS}] / B\,[\text{GB/s}]
\]
in ops/byte, where $P$ is FP16 Tensor Core peak throughput and $B$ is device memory bandwidth. For representative AI-RAN accelerators (DGX Spark, GH200, and L40S), this ridge point lies between 246 and 458~ops/byte.

We then estimate the arithmetic intensity of a representative 5G NR LDPC configuration (BG1, lifting factor $Z{=}128$, Offset Min-Sum, 10 iterations) by modeling the two dominant kernel types at edge granularity (FP16, 2~B per message):
\begin{itemize}
	\item \emph{Check-node (CN) update}: 10~ops/edge over 4~B/edge $\Rightarrow$ $\eta_{\mathrm{CN}} \approx 2.5$~ops/byte.
	\item \emph{Variable-node (VN) update}: 2~ops/edge over 10~B/edge $\Rightarrow$ $\eta_{\mathrm{VN}} \approx 0.2$~ops/byte.
\end{itemize}
BG1 has 316 non-zero entries in its base graph ($E = 316Z = 40{,}448$ edges/block). Aggregating over $I{=}10$ iterations yields total traffic of ${\approx}5.7$~MB and ${\approx}4.9{\times}10^6$~ops, which gives an overall intensity of
\begin{equation}
	\eta_\text{LDPC} \approx \frac{4.9\times10^6}{5.7\times10^6} \approx 0.9~\text{ops/byte},
\end{equation}
This is two to three orders of magnitude below the device ridge points ($\eta^* \in [246, 458]$~ops/byte), confirming that LDPC decoding is decisively memory-bandwidth-bound on these accelerators.

\subsection{Effect of SM Reservation on GPU Utilization}
\label{app:sm-enveloping-effect}

\begin{figure}[t]
	\centering
	\includegraphics[width=\linewidth]{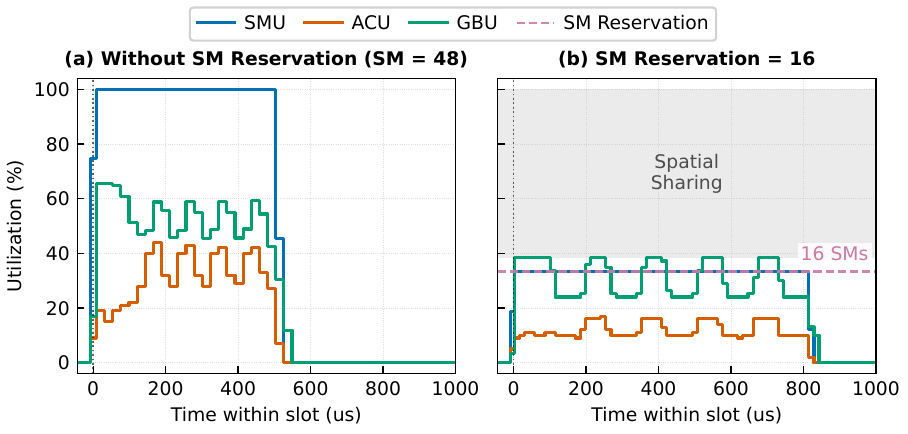}
	\caption{Per-slot GPU utilization (SMU, ACU, GBU) without and with SM reservation: the gray region shows headroom for spatial sharing.}
	\label{fig:sm-cap-utilization}
\end{figure}

\cref{fig:sm-cap-utilization} compares per-slot utilization on a 48-SM GPU (MCS~15, 162~PRBs) with and without SM reservation. The main takeaway is that SM reservation trades a modest latency increase for substantially more stable and shareable headroom.

\mypar{SM reservation creates headroom in all three dimensions} Without an SM reservation (\cref{fig:sm-cap-utilization}a), LDPC decoding bursts across all 48~SMs for ${\sim}$550\,$\mu$s (SMU ${\sim}$100\,\%, GBU ${\sim}$65\,\%, ACU ${\sim}$40\,\%), after which the GPU sits idle for the remaining ${\sim}$450\,$\mu$s. With a 16-SM reservation (\cref{fig:sm-cap-utilization}b), the same work is spread over a longer interval: SMU stabilizes at ${\sim}$33\,\% for ${\sim}$820\,$\mu$s, and the freed resources along all three dimensions become available for spatial sharing with co-located workloads.

\gls{gbu} drops because fewer SMs issue concurrent memory requests, spreading the same traffic over a longer window. \gls{acu} also drops because fewer SMs are active, but in our measurements the per-SM \gls{acu} within the active SM \emph{increases}: denser thread-block packing gives the warp scheduler more warps to interleave compute with memory stalls~\cite{volkov2016latency,cuda-prog-guide}.

\mypar{3$\times$ fewer SMs incur only 1.5$\times$ slowdown} This headroom comes at a relatively small cost in completion time. Reducing the SM allocation by $3\times$ (from 48 to 16~SMs) increases decoding latency by only $1.5\times$ (${\sim}$550\,$\mu$s to ${\sim}$820\,$\mu$s), because fewer SMs each host more resident warps, improving per-SM occupancy and hiding memory stalls more effectively.

\section{Compute-Aware Scheduler Details}
\label{app:ran-ctrl}

This appendix elaborates on \S\ref{sec:ran-control}. It first explains the lookup table that maps RAN configurations to SM demand, then presents the full pseudocode of the compute-aware UL scheduler, and finally summarizes the parameters that govern its responsiveness and stability.

\subsection{Lookup Table}
\label{app:lookup-table}

The lookup table captures the relationship between RAN configuration and SM demand. It is indexed by (MCS, PRB count, SM count) and records whether LDPC decoding meets the RAN-slot deadline for each triple. During scheduling, the controller uses it in two directions: PRB budgeting asks how many PRBs a given SM budget can support, while SM adjustment asks how many SMs the RAN requires for a given PRB demand. These two accesses are exposed as \texttt{LUT\_PRB} and \texttt{LUT\_SM}, which are different projections of the same table:

\begin{tightlist}
	\item \texttt{LUT\_PRB(mcs, sm\_count)}: fixes the MCS index and SM count, then returns the maximum PRB count whose decoding meets the RAN-slot deadline.
	\item \texttt{LUT\_SM(mcs, num\_prbs)}: fixes the MCS index and PRB count, then returns the minimum SM count that meets the RAN-slot deadline.
\end{tightlist}

We construct the table in two steps. First, we profile standalone LDPC decoding with the OAI \texttt{ulsim} tool: for each (MCS, PRB, SM count) triple, decoding is benchmarked on the target GPU with the SM reservation applied, and the P99 latency is checked against a conservative 800\,$\mu$s RAN-slot decoding budget. Second, we calibrate the table against end-to-end OAI runs with \texttt{rfsim} under synthetic extreme channel conditions and workloads, ensuring that the resulting budgets introduce no additional HARQ deadline violations beyond the baseline scheduler. For (MCS, PRB) pairs between profiled sample points, we apply ceiling interpolation in both dimensions so every query remains conservative by construction.

\subsection{Pseudocode}
\label{app:ran-ctrl-algo}

Algorithm~\ref{alg:compute-aware} instantiates the per-slot control loop described in~\cref{fig:ran-control}. It takes the standard PUSCH scheduler input together with persistent SM state, then wraps the OAI proportional-fair UL scheduler with three scheduler-specific mechanisms: PRB budgeting, adaptive SM adjustment, and iteration control. The output is a set of UL PRB grants together with the current-slot SM budget, which is retained as prior state for the next slot; its complement is current-slot spare-SM capacity.

The control flow consists of four stages: (a)~query the lookup table to obtain a provisional PRB budget, (a$'$)~run standard PUSCH scheduling within that budget, (b)~adjust the target SM count based on observed backlog and recent history, and (c)~recompute the PRB allowance under the updated state, looping whenever the current allocation is either above that allowance or more than a small tolerance below it.

\renewcommand{\algorithmiccomment}[1]{\hfill $\triangleright$ \textit{#1}}
\newcommand{\SECTIONHEAD}[1]{\STATE \textit{//~\textbf{#1}}}

\begin{algorithm*}[t]
	\caption{Compute-Aware UL Scheduler: per-slot pipeline}
	\label{alg:compute-aware}
	\footnotesize
	\setlength{\baselineskip}{9pt}
	\begin{algorithmic}[1]
		\REQUIRE Scheduler input for standard PUSCH scheduling
		\REQUIRE Persistent SM history state (carried across slots):\\
		$\textit{sm\_prev}$: prev-slot SM count;\quad
		$\textit{mcs\_prev}$: prev-slot weighted-avg MCS;\\
		$\textit{slots\_since\_under}$: underutilization counter;\quad
		$\textit{slots\_since\_update}$: slots since last SM change
		\ENSURE  PUSCH grants within current-slot SM compute budget; SM state carried to next slot
		\SECTIONHEAD{Entry-point initialisation}
		\STATE $\textit{tgt\_sm} \gets \textit{sm\_prev}$;
		$\textit{mcs\_est} \gets \textit{mcs\_prev}$;
		$\textit{iter} \gets 0$;
		$\textit{high\_demand\_seen} \gets \mathbf{false}$
		\SECTIONHEAD{Iterative scheduling pipeline (\cref{fig:ran-control}, steps a--c)}
		\STATE \textsc{PipelineEntry}: $\textit{iter} \gets \textit{iter} + 1$
		\SECTIONHEAD{Step (a): PRB Budgeting}
		\STATE $\textit{prbs\_avail} \gets \texttt{LUT\_PRB}(\textit{mcs\_est},\;\textit{tgt\_sm})$ \COMMENT{max PRBs decodable within RAN-slot deadline}
		\SECTIONHEAD{Step (a$'$): Standard PUSCH Scheduling within $\textit{prbs\_avail}$}
		\STATE Run standard PF scheduling capped by $\textit{prbs\_avail}$
		\STATE Record $\textit{prbs\_alloc} \gets \sum_u \textit{rbSize}_u$,\; $\textit{sum\_backlog} \gets \sum_u B_u$
		\STATE $\textit{mcs\_curr} \gets {\sum_{u}\; \textit{mcs}_u \cdot \textit{rbSize}_u}\;/\;{\textit{prbs\_alloc}}$
		\SECTIONHEAD{Step (b): Adaptive SM Count Adjustment}
		\STATE \textit{// Ramp-Up: immediate increase to drain backlog}
		\IF{$\exists\;u$ with $B_u > \textit{BACKLOG\_THR}$
			\textbf{and} the PRB budget is exhausted or residual backlog exceeds threshold}
		\STATE $\textit{est\_prbs} \gets \max\!\bigl(\textit{prbs\_alloc} + \textit{min\_rb},\; \texttt{est\_prbs\_from\_backlog}(\textit{sum\_backlog})\bigr)$
		\STATE $\textit{tgt\_sm} \gets \texttt{LUT\_SM}(\textit{mcs\_est},\; \textit{est\_prbs})$;\; $\textit{high\_demand\_seen} \gets \mathbf{true}$
		\ENDIF
		\STATE \textit{// Fast Exit: immediate down-sizing when slot is idle}
		\IF{\textbf{not} $\textit{high\_demand\_seen}$ \textbf{and} $\textit{prbs\_alloc}=0$ \textbf{and} $\textit{tgt\_sm}>\textit{HIGH\_TIER\_CAP}$}
		\STATE $\textit{exit\_sm} \gets \texttt{LUT\_SM}(\textit{mcs\_est},\; \texttt{est\_prbs\_from\_backlog}(\textit{sum\_backlog}))$
		\IF{$\textit{exit\_sm} \le \textit{HIGH\_TIER\_CAP}$}
		\STATE $\textit{tgt\_sm} \gets \max\!\bigl(\textit{tgt\_sm} - \textit{STEP},\; \max(\textit{exit\_sm},\;\textit{SM\_MIN})\bigr)$ \COMMENT{step down one tier}
		\ENDIF
		\ENDIF
		\STATE \textit{// Hold or Ramp-Down: hysteresis-based conservative step-down}
		\IF{$\textit{tgt\_sm} \neq \textit{sm\_prev}$}
		\STATE $\textit{slots\_since\_under} \gets 0$;\; $\textit{slots\_since\_update} \gets 0$
		\ELSE
		\STATE $\textit{slots\_since\_update} \mathrel{+}= 1$
		\IF{$\textit{sm\_prev} > \textit{SM\_MIN}$ \textbf{and} $\texttt{LUT\_PRB}(\textit{mcs\_curr},\; \textit{sm\_prev} - \textit{STEP}) \ge \textit{prbs\_alloc}$}
		\STATE $\textit{slots\_since\_under} \mathrel{+}= 1$ \COMMENT{one fewer tier would have sufficed}
		\ELSE
		\STATE $\textit{slots\_since\_under} \gets 0$
		\ENDIF
		\ENDIF
		\IF{$\textit{slots\_since\_under} \ge \textit{DECR\_THR}$}
		\STATE $\textit{desired\_sm} \gets \max\!\bigl(\texttt{LUT\_SM}(\textit{mcs\_est},\; \texttt{est\_prbs\_from\_backlog}(\textit{sum\_backlog})),\;\textit{SM\_MIN}\bigr)$
		\STATE $\textit{tgt\_sm} \gets \max\!\bigl(\textit{sm\_prev} - \textit{STEP},\; \textit{desired\_sm}\bigr)$;\; $\textit{slots\_since\_under} \gets 0$
		\ENDIF
		\SECTIONHEAD{Step (c): Iteration Control}
		\STATE $P_{\max} \gets \texttt{LUT\_PRB}\!\bigl(\textit{mcs\_curr},\; \textit{tgt\_sm}\bigr)$ \COMMENT{recompute with true MCS and updated SM}
		\IF{$\textit{prbs\_alloc} \le P_{\max}$ \textbf{and} $P_{\max} - \textit{prbs\_alloc} \le \textit{PRB\_TOL}$}
		\STATE \textsc{Success}: emit grants
		\ELSIF{$\textit{iter} < \textit{MAX\_ITER}$}
		\STATE $\textit{mcs\_est} \gets \textit{mcs\_curr}$;\; \textbf{goto} \textsc{PipelineEntry}
		\ELSIF{$\textit{prbs\_alloc} > P_{\max}$}
		\STATE \textsc{Safeguard}: force $\textit{tgt\_sm} \gets \texttt{LUT\_SM}(\textit{mcs\_curr},\; \textit{prbs\_alloc})$ \COMMENT{ramp-up to guarantee forward progress}
		\ELSE
		\STATE \textsc{Conservative Exit}: emit grants \COMMENT{safe but under-allocated after hitting the iteration bound}
		\ENDIF
		\SECTIONHEAD{Carry state to next slot}
		\STATE $\textit{sm\_prev} \gets \textit{tgt\_sm}$ \COMMENT{retain current-slot SM count as next-slot prior}
		\IF{$\textit{mcs\_curr} > 0$}
		\STATE $\textit{mcs\_prev} \gets \textit{mcs\_curr}$
		\ENDIF
		\STATE \textbf{Output:} final UL PRB grants and $\textit{tgt\_sm}$ for this slot
	\end{algorithmic}
\end{algorithm*}

\subsection{Parameters}
\label{app:ran-ctrl-params}

Table~\ref{tab:ran-ctrl-params} lists the tunable parameters used by the scheduler. These parameters control four things: the SM quantization granularity, how aggressively the scheduler ramps up in response to backlog, how cautiously it ramps down after demand subsides, and how tightly iteration control tracks the recomputed PRB allowance. We choose their values through end-to-end profiling on the DGX Spark testbed across different traffic scenarios. The tuning process sweeps each parameter while measuring SM reservation stability (transition count), HARQ deadline-met rate, and UE-level KPIs (throughput, delay, and loss), then selects values that improve the reservation stability without degrading RAN performance.

\begin{table}[t]
	\centering
	\floatfont
	\caption{Compute-aware scheduler parameters.}
	\label{tab:ran-ctrl-params}
	\begin{tabular}{llr}
		\toprule
		\textbf{Symbol}       & \textbf{Description}             & \textbf{Value} \\
		\midrule
		\textit{SM\_MIN}      & Minimum SM count (floor)         & 8 SMs          \\
		\textit{SM\_MAX}      & Maximum SM count (all SMs)       & 48 SMs         \\
		\textit{STEP}         & SM increment/decrement step      & 8 SMs          \\
		\textit{BACKLOG\_THR} & Per-UE backlog ramp-up threshold & 64 KB          \\
		\textit{DECR\_THR}    & Ramp-down patience (slots)       & 30 slots       \\
		\textit{PRB\_TOL}     & PRB slack tolerance in step (c)  & 10 PRBs        \\
		\textit{MAX\_ITER}    & Pipeline iteration bound         & 5              \\
		\bottomrule
	\end{tabular}
\end{table}

\mypar{SM\_MIN and STEP} define the scheduler's SM control granularity. The SM count is quantized in steps of 8, matching the CUDA Green Context co-scheduling granularity observed on the DGX Spark. Setting the minimum to 8~SMs ensures that even the lightest RAN workload can still be decoded within the RAN-slot deadline.

\mypar{BACKLOG\_THR} governs when the scheduler switches from smoothing to draining. We set it to 64\,KB, approximately half of a full-bandwidth transport block at moderate MCS. A smaller value triggers ramp-up more eagerly, improving UE latency but causing more SM transitions; a larger value absorbs more traffic bursts at a constant SM tier, reducing transitions at the cost of slightly higher buffer occupancy.

\mypar{DECR\_THR} governs how much evidence of over-provisioning is required before releasing SMs. We set the ramp-down patience to 30 consecutive RAN slots (${\sim}$15\,ms at 30\,kHz SCS). The scheduler therefore steps down only after sustained over-provisioning, preventing premature SM release from fragmenting the SM reservation. This value is chosen to exceed the typical burst duration observed in the RAN.

\mypar{PRB\_TOL} bounds how much PRB under-allocation is tolerated in iteration control. We set it to 10~PRBs, so the scheduler accepts small slack gaps without another pass.

\mypar{MAX\_ITER} bounds the iteration loop. We cap the pipeline at 5 iterations per slot. In practice, it converges in 1--3 iterations in $>$99\% of slots; this bound is only a safeguard against non-convergence under extreme channel conditions or workloads.

\section{Two-Level Elastic Training Details}
\label{app:training}

\subsection{Scheduling Mechanism Details}
\label{app:scheduling-mechanism-details}

This subsection provides implementation details for the two levels defined in \S\ref{sec:training-strategy}: inter-site scheduling and intra-site scheduling. Within the inter-site level, \gls{air} is the fast path for transient fluctuations and \gls{mlr} is the slow path for persistent imbalance.

\mypar{Coordinator metadata}
The coordinator maintains three types of runtime state:
(i)~\emph{executor performance state}, recent throughput and latency observations used to predict completion times for micro-batch stage tasks and decide when to trigger load rebalancing;
(ii)~\emph{sample-stage commit state}, micro-batch and per-sample identifiers, attempt tokens, and commit records that provide at-most-once acceptance even when work is rerouted or speculatively re-executed; and
(iii)~\emph{deferral statistics}, the fraction of candidate micro-batches deferred in each training step, which distinguishes transient overload from persistent imbalance and drives the \gls{mlr} slow-path trigger.

\mypar{Coordinator deployment}
In RAN deployments, the coordinator runs at the core to exploit strong site-to-core links; in peer-to-peer clusters, the same logic can be hosted on a worker with replicated metadata storage.

\mypar{Rebalancing protocol}
When \gls{mlr} fires, the slow path shifts the pipeline-parallel partition boundary so that persistently overloaded stages shed layers to less-loaded sites, rather than migrating the full runtime data path.
Only the persistent state for layers that cross the boundary (parameters and optimizer state) is transferred; transient activations are regenerated under the new layout.
The \textsc{Prepare} phase drains work that crosses the changing boundary at a training-step barrier. The subsequent \textsc{Copy} and \textsc{Commit} phases transfer the moved layers' state and activate the new layout while work on unaffected stages continues, so the seconds-long transfer does not become a seconds-long exposed stall (\cref{tab:reshard-breakdown-ts}).
The complete \textsc{Prepare/Copy/Commit} protocol ensures that layout transitions are consistent and do not corrupt gradients.

\mypar{GEMM-tile admission and profiling}
To avoid runtime jitter, \sys pre-allocates GPU GEMM-tile buffers and stream state during initialization, eliminating online allocator overhead (100--300\,$\mu$s under load) from the GEMM-tile admission path.
Before deployment, \sys profiles \gls{gemm} performance across tile sizes and spare-SM budgets to build a lookup table that maps tile size $\times$ spare-SM budget to predicted completion time, so runtime GEMM-tile admission can enforce the RAN-slot deadline and bandwidth constraints without per-decision profiling.

\mypar{Fault tolerance}
\sys handles crash-stop site failures through the same mechanisms used for capacity fluctuations.
When a site becomes unreachable, the coordinator detects the failure via lease timeout, increments the task epoch for each affected micro-batch stage task, revokes the failed attempts, and reassigns pending micro-batch stage tasks to surviving replicas through \gls{air}'s min-cost assignment (Appendix \ref{app:scheduling-formalism}).
Fenced attempts guarantee at-most-once acceptance: any delayed output from the failed site carries a stale task epoch and is rejected at commit (Appendix \ref{app:air-proofs}).
For persistent site loss, \gls{mlr} rebalances the model layout to exclude the unavailable site, transferring the affected partition's parameters and optimizer state.
We do not address Byzantine faults, correlated failures across a majority of sites, or coordinator crash recovery; the coordinator is assumed to be a reliable service, deployable with standard replication techniques if higher availability is required.

\begin{figure}[t]
	\centering
	\includegraphics[width=.85\linewidth]{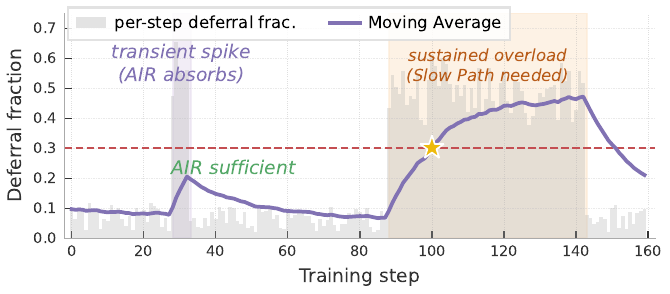}
	\caption{MLR trigger based on AIR deferrals. \markReshard{} marks the rebalancing trigger.}
	\label{fig:cross-site-deferral-trigger}
\end{figure}

\mypar{MLR example}
\cref{fig:cross-site-deferral-trigger} illustrates how MLR distinguishes transient overload from persistent imbalance.
Initially, deferrals remain low and the moving average stays well below the threshold, so the \gls{air} fast path alone is sufficient.
Later, deferrals remain high over many consecutive training steps; the moving average climbs above the threshold and stays there, at which point \gls{mlr} adjusts the model-parallel layout.
\gls{mlr} only reacts when overload is sustained, avoiding expensive rebalancing (which requires state migration) for short-lived spikes that \gls{air} can handle on its own.

\subsection{Correctness Guarantees}
\label{app:correctness}

Correctness for \sys is defined relative to a reference run that trains the same model with standard synchronous data-parallel \gls{sgd} on a fixed cluster. Dynamic rerouting, deferral, and model-layout rebalancing should not change which samples are counted or how their gradients are combined; they should only change where and when the work runs.

\mypar{At-most-once, version-valid sample-stage acceptance}
Each micro-batch stage task carries a metadata tuple that includes its training step $t$, micro-batch identifier $b$, sample-identifier set $U_b$, stage $s$, and a fenced attempt token. The coordinator maintains one commit record per sample-stage key $(t,u,s)$ for $u\in U_b$. AIR transfers each micro-batch stage task as a bundle. A normal \gls{air} transfer revokes the old attempt before authorizing a new one; the speculative retry in \cref{fig:air} may temporarily execute both copies, but an atomic open-to-committed transition accepts only the first valid result for each key. Tokens also carry a task epoch, model version, and layout epoch, so late outputs from revoked attempts or old layouts are rejected rather than applied twice. Thus each sample-stage result is accepted at most once per training step and no stale gradient is applied.

\mypar{Per-sample gradient correctness with varying participation}
Because the \gls{air} fast path may carry infeasible micro-batch stage tasks into the next training step, replicas can complete different sample counts $n_{g,t}$ in a training step and the effective global batch size $B_t=\sum_g n_{g,t}$ can vary over time. \sys aggregates gradients using the weighted rule in Eq.~\eqref{eq:weighted-aggregation}, which uses the actual sample counts $n_{g,t}$ from each replica instead of assuming fixed sample quotas. This produces the exact per-sample average over the set of samples accepted in training step $t$, matching what a reference synchronous run would compute on that same sample set. The debt update in Eq.~\eqref{eq:debt-rebalance} then adjusts future sample quotas when a replica under-contributes for many training steps, which keeps long-run participation balanced without changing the per-training-step estimator. Appendix \ref{app:air-proofs} gives the convergence argument under this varying effective batch size.

\mypar{Bounded latency and eventual progress from fast- and slow-path inter-site adaptation}
\gls{air} derives a per-training-step admission deadline from the SM reservation published by RAN Control and admits only micro-batch stage tasks whose predicted finish time does not exceed this deadline. Micro-batch stage tasks that would miss the deadline are rerouted when another equivalent-stage executor is feasible and otherwise remain unowned until admission in the next training step. When persistent capacity loss keeps the deferral rate high, \gls{mlr} adjusts pipeline boundaries and sample quotas until the new layout again satisfies the timing constraint. Thus \gls{air} handles transient inter-site variation, \gls{mlr} handles persistent inter-site imbalance, and the intra-site scheduler independently absorbs RAN-slot-level changes through GEMM-tile admission.

\mypar{Summary of guarantees}
\gls{air} rerouting and deferral change where and when micro-batch stage tasks execute, but not how accepted gradients are formed or which model and layout version they update: weighted aggregation is unbiased when admission, deferral, and rerouting do not depend on sample content; the per-key atomic commit gate prevents duplicate contributions during rerouting or speculative re-execution; the debt update keeps long-run contribution skew bounded for $0<\kappa<1$; and during \gls{mlr}, the barrier-aligned \textsc{Prepare/Copy/Commit} protocol binds every accepted completion to one layout epoch. Appendix \ref{app:air-proofs} gives the corresponding derivations and protocol arguments.

\subsection{Scheduling Formalism}
\label{app:scheduling-formalism}

This subsection defines the latency, aggregation, and debt notation that formalizes the scheduling behavior described in \S\ref{sec:training-strategy}.

\gls{air} enforces a per-training-step admission deadline defined by the nominal barrier deadline $T_{\text{bar}}$ with bounded elastic slack $\Delta_t\le\Delta_{\max}$. Admitting only micro-batch stage tasks that finish by this deadline yields the training-step-latency bound
\begin{equation}
	L_t \le T_{\text{bar}}+\Delta_{\max}+\Delta_c,
	\label{eq:latency-bound}
\end{equation}
where $L_t$ is observed training-step latency, $\Delta_{\max}$ is the configured maximum elastic slack, and $\Delta_c$ is the duration of one inter-site AIR control interval. RAN Control exports the SM reservation; the training scheduler derives these timing quantities from that reservation and recent completion-time errors.

After rerouting/deferral, replica $g$ may complete $n_{g,t}$ samples at training step $t$, with summed local gradient contribution $G_{g,t}$. The tree-reduce through the hub computes
\begin{equation}
	\bar{g}_t = \frac{\sum_{g=1}^{G} G_{g,t}}{\sum_{g=1}^{G} n_{g,t}}.
	\label{eq:weighted-aggregation}
\end{equation}
Here $\bar{g}_t$ is the global per-sample average over all uniquely accepted samples in training step $t$, $B_t=\sum_g n_{g,t}$ is the realized effective global batch size, and $G$ is the number of participating replicas in that step.

To correct persistent contribution skew, \sys maintains per-replica debt state $d_g$ and updates next-step sample quota $q_g$ using
\begin{align}
	d_g(t{+}1) & = d_g(t) + \left(n^{\text{target}} - n_{g,t}\right),                       \\
	q_g(t{+}1) & = \operatorname{clip}\!\left(q_g^{\text{base}} + \kappa d_g(t{+}1)\right).
	\label{eq:debt-rebalance}
\end{align}
where $n^{\text{target}}$ is the target per-replica sample count, $q_g^{\text{base}}$ is the baseline sample quota from placement, and $\kappa$ is the debt-correction gain.

\subsection{AIR Correctness Arguments}
\label{app:air-proofs}

This subsection provides the derivations and protocol arguments behind \sys's aggregation, commit, debt, and layout-transition semantics.

\mypar{Unbiased weighted aggregation}
For training step $t$, let $\mathcal{S}_t$ be the set of accepted samples after rerouting and deduplication. For each sample $i\in\mathcal{S}_t$, let $g_i(w_t)$ denote its per-sample gradient at synchronized weight $w_t$.
Replica $g$ contributes
\begin{equation}
	G_{g,t} = \sum_{i\in\mathcal{S}_{g,t}} g_i(w_t), \qquad n_{g,t}=|\mathcal{S}_{g,t}|,
	\label{eq:gradient-local-sum}
\end{equation}
with disjoint partition $\mathcal{S}_t=\bigsqcup_g \mathcal{S}_{g,t}$ due to at-most-once acceptance.
The estimator in Eq.~\eqref{eq:weighted-aggregation} can be rewritten as
\begin{equation}
	\bar{g}_t = \frac{\sum_g G_{g,t}}{\sum_g n_{g,t}}
	= \frac{1}{|\mathcal{S}_t|}\sum_{i\in\mathcal{S}_t} g_i(w_t).
	\label{eq:gradient-global-avg}
\end{equation}
Hence it is exactly the sample average over all accepted samples. Under standard random sampling of the global batch and the assumption that admission, deferral, and rerouting decisions are independent of sample content conditioned on system state, $\mathbb{E}[\bar{g}_t\mid w_t]$ equals the gradient of the accepted-sample objective. The weighted aggregation rule is therefore unbiased under this content-independent scheduling assumption.

\mypar{At-most-once acceptance under fenced attempts}
Each sample-stage key is $\chi=(t,u,s)$ with a commit record $c_\chi\in\{\textsc{open},\textsc{committed}\}$ and one or more authorized attempt tokens $\tau_a$ tied to the current task epoch, model version, and layout epoch. A normal ownership transfer increments the task epoch and revokes the old token; speculative re-execution may authorize an additional token in the same task epoch. Every result must first pass token and version validation, then atomically attempt
\begin{equation}
	c_\chi:\textsc{open} \xrightarrow{\mathrm{CAS}} \textsc{committed}.
	\label{eq:commit-cas}
\end{equation}

\emph{Safety (at-most-once sample-stage acceptance):}
Suppose two results for the same key $\chi$ are accepted. Both must successfully change $c_\chi$ from \textsc{open} to \textsc{committed}, but the atomic \gls{cas} permits only one such transition. This remains true when two speculative copies have valid tokens; the slower copy observes an already committed record and is discarded. Results from earlier task or layout epochs fail validation before reaching the commit gate.

\emph{Liveness (progress with failures):}
If an executor fails, lease timeout triggers reassignment with a higher task epoch. Any delayed old-attempt output is stale and rejected. Therefore, if a healthy executor eventually obtains a valid attempt token and completes, one result can commit while all later results are rejected.

\mypar{Bounded contribution debt}
Define debt dynamics (Eq.~\eqref{eq:debt-rebalance}) and let realized count be
\begin{equation}
	n_{g,t}=n^{\text{target}}+\kappa d_g(t)+\xi_{g,t},
	\label{eq:debt-realized-count}
\end{equation}
where $\xi_{g,t}$ captures bounded actuation/scheduling noise, $|\xi_{g,t}|\le \Xi$.
Then
\begin{equation}
	d_g(t{+}1)=(1-\kappa)d_g(t)-\xi_{g,t}.
	\label{eq:debt-linear-dyn}
\end{equation}
For $0<\kappa<1$, this is a stable linear system with bounded disturbance. Recursing gives
\begin{equation}
	|d_g(t)| \le (1-\kappa)^t|d_g(0)| + \frac{\Xi}{\kappa}.
	\label{eq:debt-bound}
\end{equation}
Thus long-run skew is bounded by $O(\Xi/\kappa)$ and decays geometrically when noise is small.

\mypar{Layout-transition consistency}
MLR cutover begins at a training-step barrier $t^\star$, where the pipeline is quiescent at the changing boundary. In the \emph{PREPARE} phase, \sys freezes \gls{air} assignments whose path would cross the moved layers, drains in-flight work under the old layout, and stops admitting new work at the affected stages. In the \emph{COPY} phase, it transfers parameters and optimizer state $S_\ell$ for the moved layers and verifies checksums without changing attempt metadata; work that does not traverse the changing boundary can continue. In the \emph{COMMIT} phase, it atomically publishes the new boundary vector, rebuilds equivalent-stage pools $\{\mathcal{E}_s\}$, and increments the monotonic layout epoch $r$ in the coordinator ledger. After COMMIT, any new micro-batch stage task that enters the affected stages carries the updated layout epoch in its metadata $\langle t,b,U_b,s,e,v,\tau_a,r\rangle$, where $e$ is the task epoch, $v$ is the model version, and $r$ is the layout epoch, and the coordinator accepts a completion for key $\chi=(t,u,s)$ for each $u\in U_b$ only if its task epoch $e$, model version $v$, attempt token $\tau_a$, and layout epoch $r$ are current. Hence any accepted completion for $\chi$ is executed entirely under a single pipeline layout.

Extending the fenced-attempt validation to include $r$ preserves at-most-once acceptance across boundary moves. Suppose two distinct results for the same key $\chi$ were both accepted around an MLR event. If they carry the same current task, model, and layout versions, they still compete at the single open-to-committed \gls{cas}, which accepts only one. If they carry different layout epochs $r<r'$, validation accepts only the current layout epoch, so the stale result is rejected before the commit gate. Thus at most one result for each $\chi$ is accepted, and no commit contains mixed or stale layout information. Combined with the liveness condition above, this yields eventual exactly-once sample contribution.

\mypar{Protocol invariant summary}
Taken together, these arguments establish one end-to-end acceptance invariant per sample-stage key $\chi=(t,u,s)$: (i) metadata-gated validation on $\langle t,b,U_b,s,e,v,\tau_a,r\rangle$ and $u\in U_b$ enforces task-attempt, model-version, and layout compatibility; (ii) the per-key open-to-committed \gls{cas} allows at most one accepted result even under speculative re-execution; (iii) PREPARE/COPY/COMMIT performs an atomic cutover at training-step barrier $t^\star$, so each accepted completion is bound to one layout epoch $r$; and (iv) outputs with stale $(e,v,\tau_a)$ or stale $r$ are rejected, while reclamation is deferred until drain completion.



\section{Evaluation Details}
\label{app:eval}

\subsection{RAN-Centric Spare Compute Controller Evaluation}
\label{app:ran-eval-methodology}

\subsubsection{RAN configuration}
\label{app:ran-config}
The gNB runs 5G NR standalone (SA) on band~n78 (3.5\,GHz TDD) with 106~PRBs at 30\,kHz subcarrier spacing (numerology~1), corresponding to a 40\,MHz carrier. The TDD pattern uses a 5\,ms periodicity with 7~DL slots, 1~flexible slot, and 2~UL slots (DDDDDDDSUU), yielding a ${\sim}$20\,\% UL duty cycle. All experiments run on an NVIDIA DGX Spark with 48~SMs.

\subsubsection{Channel condition replay}
\label{app:channel-replay}
Static AWGN channels represent the primary fidelity gap between RF-simulated and OTA experiments. We close this gap by replaying per-UE uplink channel measurements from the TRACTOR dataset~\cite{tractor} into OAI \texttt{RFSIM}'s channel model. At each trace timestamp, the recorded UL RSSI and SINR are translated into the simulator's path-loss and noise-power controls:
\begin{align}
	\texttt{ploss\_dB}_{u,k}        & = P_{\mathrm{ref}} - \mathrm{RSSI}_{u,k}, \\
	\texttt{noise\_power\_dB}_{u,k} & = N_{\mathrm{off}} - \mathrm{SINR}_{u,k},
\end{align}
where $P_{\mathrm{ref}}$ and $N_{\mathrm{off}}$ are fixed calibration offsets. Updates are sent via OAI's telnet interface and synchronized with traffic generators using Unix wall-clock timestamps, so each UE observes the same time-varying channel trajectory as the original trace.

\subsubsection{Application traffic replay}
\label{app:traffic-replay}
\hfill\\
\mypar{TRACTOR application trace replay}
We use production O-RAN traffic from the TRACTOR 5G dataset~\cite{tractor}, selecting two four-UE scenarios: Scenario~1 (referred to as \textit{Multi-UE/Trial2/multi4} in the dataset) and Scenario~2 (referred to as \textit{Multi-UE/Trial3/multi4} in the dataset). Per-UE traffic profiles and channel conditions are listed in Table~\ref{tab:tractor-datasets}. For the RAN-control evaluation, traces are replayed with the original packet timing over a 600\,s window; shorter UE traces are looped to ensure all four UEs remain active.

\mypar{D-ITG stress traffic}
For UE-level KPI evaluation, we use D-ITG~\cite{ditg} to generate synthetic bursty UDP traffic that exercises more extreme conditions than production traces. Each UE runs a steady 800\,kbps flow plus a 20\,s burst at 3.6\,Mbps, so the aggregate peak load (${\sim}$13.2\,Mbps across three UEs) saturates the uplink.

\subsubsection{HARQ deadline-met rate}
\label{app:harq-deadline}
We instrument the OAI MAC scheduler to record the expected frame and slot
for UL HARQ feedback. A waiting HARQ process is counted as overdue when its
feedback remains missing beyond the configured grace window; the same check
applies when feedback arrives for a different pending process. The scheduler
exports cumulative overdue events and initial UL transport-block transmissions.
We use differences of these counters over the measurement window. The
normalized HARQ deadline-met metric is
$1-(\text{overdue events}/\text{initial UL TBs})$; for live co-location we
report the underlying event rate per 1000 initial UL TBs. This MAC-level
metric is separate from the fraction of LDPC decoder records exceeding a
1\,ms processing budget.

\begin{table}[t]
	\centering
	\caption{TRACTOR dataset configurations used in RAN control evaluation.}
	\label{tab:tractor-datasets}
	\clampbox{\columnwidth}{\setlength{\tabcolsep}{3pt}
		\begin{tabular}{llllr}
			\toprule
			\textbf{Datasets} & \textbf{UE} & \textbf{Traffic profile}        & \textbf{Type} & \textbf{Avg.\ UL SINR} \\
			\midrule
			Scenario\,1       & UE\,1       & Background traffic              & mMTC          & 13.1\,dB              \\
			                  & UE\,2       & YouTube streaming (stationary)  & eMBB          & 2.3\,dB               \\
			                  & UE\,3       & Background traffic              & mMTC          & 3.3\,dB               \\
			                  & UE\,4       & YouTube streaming (stationary)  & eMBB          & 6.0\,dB               \\
			\midrule
			Scenario\,2       & UE\,1       & Netflix streaming (stationary)  & eMBB          & 20.7\,dB              \\
			                  & UE\,2       & YouTube streaming (campus walk) & eMBB          & 5.5\,dB               \\
			                  & UE\,3       & Background traffic              & mMTC          & 4.2\,dB               \\
			                  & UE\,4       & Internet browsing (stationary)  & eMBB          & 12.2\,dB              \\
			\bottomrule
		\end{tabular}}
\end{table}

\subsection{Training Evaluation Settings}
\label{app:eval-settings}

Here, rank~$i$ denotes the worker at site~$i$. \cref{tab:internode-bandwidth} provides the inter-site bandwidth and \cref{tab:internode-latency-ms} provides the inter-site latency.

\begin{table}[t]
	\centering
	\small
	\caption{Inter-site bandwidth matrix (src\_rank $\rightarrow$ dst\_rank), in Gbps. Diagonal entries are not applicable (---).}
	\label{tab:internode-bandwidth}
	\clampbox{\linewidth}{\setlength{\tabcolsep}{3pt}
		\begin{tabular}{lrrrrrrrr}
			\toprule
			\textbf{src/dst} & \textbf{rank0} & \textbf{rank1} & \textbf{rank2} & \textbf{rank3} & \textbf{rank4} & \textbf{rank5} & \textbf{rank6} & \textbf{rank7} \\
			\midrule
			rank0            & ---            & 11.0           & 0.3            & 1.0            & 0.9            & 0.9            & 1.0            & 1.0            \\
			rank1            & 20.5           & ---            & 0.3            & 1.0            & 1.0            & 0.9            & 1.0            & 1.0            \\
			rank2            & 1.1            & 1.0            & ---            & 0.9            & 0.9            & 10.6           & 1.0            & 1.0            \\
			rank3            & 1.1            & 1.0            & 0.3            & ---            & 14.6           & 0.9            & 1.0            & 1.0            \\
			rank4            & 1.1            & 1.0            & 0.3            & 11.9           & ---            & 1.0            & 1.0            & 1.0            \\
			rank5            & 1.1            & 1.0            & 10.2           & 0.9            & 0.9            & ---            & 1.0            & 1.0            \\
			rank6            & 1.0            & 0.8            & 0.3            & 0.8            & 0.8            & 0.8            & ---            & 14.5           \\
			rank7            & 1.0            & 0.9            & 0.3            & 0.8            & 0.9            & 0.8            & 17.9           & ---            \\
			\bottomrule
		\end{tabular}}
\end{table}

\begin{table}[t]
	\centering
	\small
	\caption{Inter-site latency matrix (src\_rank $\rightarrow$ dst\_rank), in ms. Diagonal entries are not applicable (---).}
	\label{tab:internode-latency-ms}
	\setlength{\tabcolsep}{4pt}
	\clampbox{\linewidth}{\begin{tabular}{lrrrrrrrr}
			\toprule
			\textbf{src/dst} & \textbf{rank0} & \textbf{rank1} & \textbf{rank2} & \textbf{rank3} & \textbf{rank4} & \textbf{rank5} & \textbf{rank6} & \textbf{rank7} \\
			\midrule
			rank0            & ---            & 0.8            & 25.9           & 26.7           & 26.7           & 25.7           & 23.7           & 23.5           \\
			rank1            & 0.8            & ---            & 25.7           & 26.5           & 26.7           & 26.5           & 23.6           & 23.7           \\
			rank2            & 25.7           & 25.8           & ---            & 28.8           & 28.8           & 0.8            & 25.7           & 25.6           \\
			rank3            & 26.4           & 26.5           & 28.6           & ---            & 30.5           & 28.7           & 26.4           & 26.6           \\
			rank4            & 26.6           & 26.6           & 28.8           & 0.9            & ---            & 28.6           & 26.6           & 26.5           \\
			rank5            & 25.8           & 25.9           & 0.9            & 28.8           & 28.8           & ---            & 25.6           & 25.7           \\
			rank6            & 23.7           & 23.6           & 25.7           & 26.5           & 26.6           & 25.8           & ---            & 0.2            \\
			rank7            & 23.5           & 23.5           & 25.6           & 26.6           & 26.7           & 25.7           & 0.2            & ---            \\
			\bottomrule
		\end{tabular}}
\end{table}

\subsection{Training Stability and Convergence}
\label{app:stability-convergence}

\mypar{Convergence semantics}
Our training-side experiments focus on throughput and efficiency, but all systems run under strict synchronous semantics: \gls{air} accepts each completion at most once, gradients are aggregated with the unbiased weighted rule over accepted samples, and no stale gradients are applied ($K{=}0$), so convergence on the completed-sample objective matches a reference synchronous run (Appendices~\ref{app:correctness}--\ref{app:air-proofs}).

\begin{figure}[t]
	\centering
	\includegraphics[width=\linewidth]{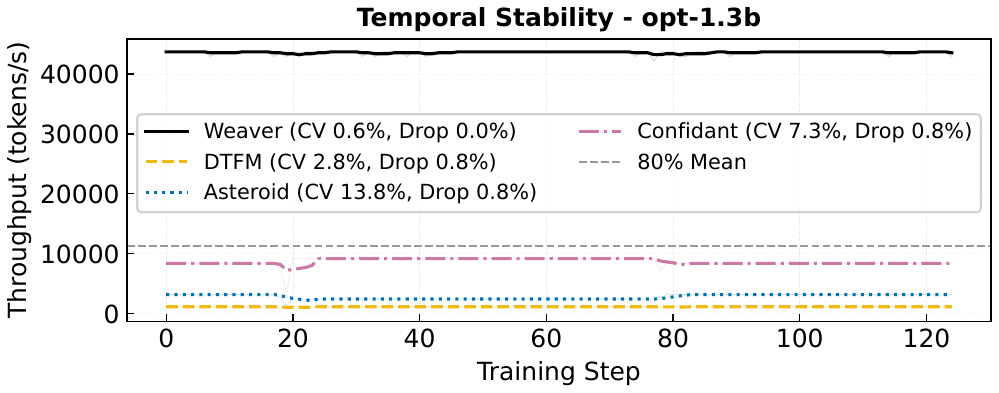}
	\caption{Step-level training throughput under identical replayed RAN SM reservations.}
	\label{fig:temporal-stability}
\end{figure}

\cref{fig:temporal-stability} shows step-level token throughput for all systems over the same replayed RAN SM reservation.
\cref{fig:convergence-opt13} supports the loss-parity argument of Appendix~\ref{app:air-proofs} with an OPT-1.3B training-loss trace under replayed RAN SM reservations: under at-most-once acceptance and weighted aggregation, the loss trajectory tracks a fixed-cluster synchronous reference.

\begin{figure}[t]
	\centering
	\includegraphics[width=0.8\linewidth]{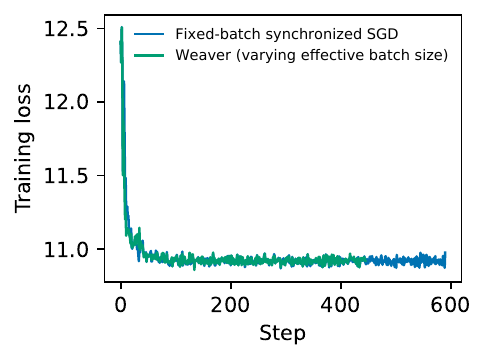}
	\caption{OPT-1.3B training-loss trace under replayed RAN SM reservations vs.\ a fixed-cluster synchronous reference.}
	\label{fig:convergence-opt13}
\end{figure}

\subsection{Inter-Site Scheduling}
\label{sec:eval-inter-scheduling}

We now evaluate how the two-level inter-site scheduler in
\S\ref{sec:two-level-scheduling} behaves under SM reservations derived from replayed RAN traces.
We feed a representative 6-minute production traffic trace through our
training simulator on OPT-13B across 8 edge sites with NVIDIA L40S
GPUs, so any difference comes only from the policies described in
\S\ref{sec:two-level-scheduling}.

For the training side, we compare \sys against the same DTFM, Asteroid,
and Confidant baselines described in~\cref{sec:eval-real-training}, which
all react at slow timescales while sharing the common oracle-MPS
substrate.

\begin{figure*}[t]
	\centering
	\includegraphics[width=\linewidth]{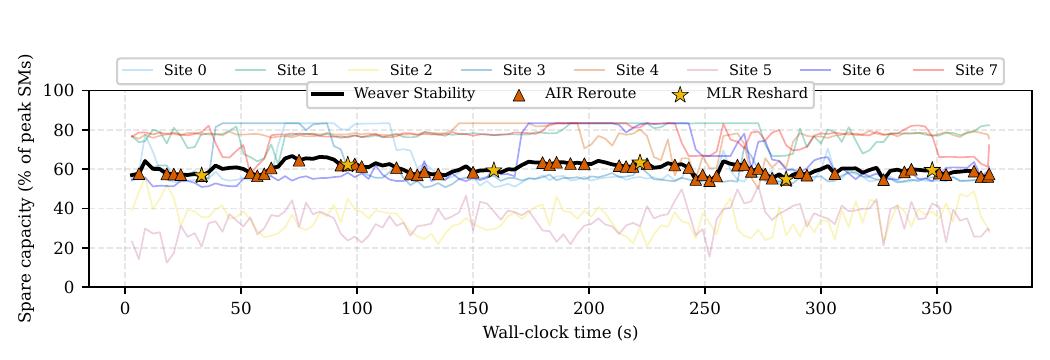}
	\caption{Timeline of inter-site scheduling over a 6-minute replay
		window under per-site SM reservations.}
	\label{fig:training-strategy-timeline}
\end{figure*}

\mypar{Fast-path AIR absorbs most capacity fluctuations}
\cref{fig:training-strategy-timeline} summarizes how Asynchronous
Inter-site Rerouting behaves when driven by SM reservations derived from real RAN traces.
Over the 6-minute window, spare compute fluctuates 80 times at the
training-step timescale (2--8\,s).
Of these events, 86.2\% are handled entirely by fast-path \gls{air} rerouting
with a median exposed stall of 66\,ms (P95: 68\,ms), and only 11 events
escalate to the slow path.
This matches the intent of the \gls{air} design in
\S\ref{sec:two-level-scheduling}: the coordinator moves work, not
state, so most transient drops in spare compute are hidden behind
normal training progress.

The runtime view in \cref{fig:training-strategy-timeline} also shows
that \gls{air} decisions occur within the current training step rather than
waiting for a global barrier.
The coordinator reassigns pending micro-batch stage tasks to equivalent-stage
executors in other data-parallel replicas, with no GPU
reconfiguration, no \gls{mps} reset, and no checkpoint replay.
As a result, the RAN-slot deadline is not endangered even when site
capacity swings quickly, which is the safety property targeted
by the inter-site scheduler in \S\ref{sec:two-level-scheduling}.

\mypar{Slow-path model layout rebalancing is rare and low-overhead}
When the SM reservation drifts in a persistent way, fast-path \gls{air} starts to
defer a growing fraction of micro-batches.
Each deferral is still within the bound enforced by the inter-site
policy, but the accumulated signal in the \gls{ewma} trigger state $z_t$
eventually crosses the MLR trigger threshold and activates slow-path \gls{mlr}.

In \cref{fig:training-strategy-timeline}, the middle panel tracks
$z_t$ and the MLR trigger threshold over time.
Across the 6-minute trace, the \gls{mlr} trigger fires 11 times, which
confirms that the deferral-based trigger in
\S\ref{sec:two-level-scheduling} reacts only when imbalance is
sustained rather than to short spikes.

To make the cost of each MLR rebalancing event concrete,
\cref{tab:reshard-breakdown-ts} decomposes the \gls{mlr} protocol into
its phases.
These numbers come from the same trace replay used above and guide the choice of rebalancing thresholds.
The PREPARE phase is the only component on the critical path, and its
528\,ms stall cost is amortized over many training steps between
MLR rebalancing events.
COPY and COMMIT mostly overlap with useful work on neighboring stages.
Taken together with the 11-event count above, this supports the design
choice in \S\ref{sec:two-level-scheduling} to keep layout changes rare
and hide most of their cost behind ongoing execution.

\begin{table}[t]
	\centering
	\floatfont
	\caption{MLR rebalancing overhead decomposition per event, averaged over 11 triggers in the 6-minute trace.}
	\label{tab:reshard-breakdown-ts}
	\begin{tabular}{lrr}
		\toprule
		Phase             & Total (ms)      & Stall (ms)     \\
		\midrule
		PREPARE (drain)   & 528.0           & 528.0          \\
		COPY (transfer)   & 4350.6          & 0.0            \\
		COMMIT (activate) & 550.0           & 0.0            \\
		\midrule
		\textbf{Total}    & \textbf{5428.6} & \textbf{528.0} \\
		\bottomrule
	\end{tabular}
\end{table}

\begin{table}[t]
	\centering
	\caption{Threshold-sensitivity analysis for inter-site scheduling under trace replay. Higher throughput and lower coefficient of variation (CoV) are better.}
	\label{tab:sensitivity}
	\clampbox{\linewidth}{
		\setlength{\tabcolsep}{3pt}
		\begin{tabular}{lccc}
			\toprule
			\textbf{Threshold}         & \textbf{Throughput} & \textbf{CoV (\%)} & \textbf{\# MLR events} \\
			\midrule
			$\theta_{\text{AIR}}$=0.05 & 21765.72 tokens/s   & 0.67              & 0                     \\
			$\theta_{\text{AIR}}$=0.1  & 21837.73  tokens/s  & 0.08              & 0                     \\
			$\theta_{\text{AIR}}$=0.2  & 21845.08 tokens/s   & 0.01              & 0                     \\
			\midrule
			$\theta_{\text{MLR}}$=0.3  & 21813.14 tokens/s   & 0.45              & 14                    \\
			$\theta_{\text{MLR}}$=0.4  & 21822.01 tokens/s   & 0.39              & 10                    \\
			$\theta_{\text{MLR}}$=0.5  & 21841.99 tokens/s   & 0.13              & 1                     \\
			\bottomrule
		\end{tabular}
	}
\end{table}

\mypar{Inter-site scheduling parameter sensitivity}
Finally, we study how the thresholds that drive AIR and MLR affect
throughput and stability under the same replayed SM reservations used above.
We sweep the AIR slack threshold $\theta_{\text{AIR}}$ and the
MLR trigger threshold $\theta_{\text{MLR}}$ while keeping all
other settings fixed (\cref{tab:sensitivity}).

Across these settings, throughput varies only slightly while the \acrshort{cov}
remains well below 1\%, which indicates stable training-step latency.
More aggressive MLR triggering increases the number of rebalancing events
but does not improve throughput in a meaningful way.
This backs up the design in \S\ref{sec:two-level-scheduling}: a
conservative trigger that lets \gls{air} handle most transients and invokes
MLR rarely is enough to keep the system both stable and efficient.

\subsection{Large-Scale Simulator Methodology}
\label{app:sim-methodology}

Our simulator extends Morphling~\cite{morphling}, a measurement-driven emulator for distributed ML that runs unmodified training scripts over fitted device performance models, with inter-site topology modeling and per-site RAN SM-reservation replay.

\mypar{Compute profiles}
Because all transformer layers are architecturally identical, we profile per-layer forward and backward latency separately for the embedding, a single transformer block, and the LM head on every testbed node, sweeping micro-batch size and sequence length (5 warm-up and 20 measured iterations per configuration.
``Running training with a reduced number of layers'' refers to exactly this: we execute truncated-depth models to obtain measured per-layer profiles rather than extrapolate from analytical FLOP counts, then reconstruct the full-model step time as embedding $+\,N_{\text{layers}}\times$block $+$ head under the chosen parallelism layout and pipeline schedule, keeping the target global batch size and sequence length unchanged.

\mypar{Communication}
We transmit the actual hidden-state and gradient tensors required by the training framework over the real links and use the measured inter-site bandwidth/latency matrices (Tables~\ref{tab:internode-bandwidth}--\ref{tab:internode-latency-ms}); the simulator additionally models per-link contention when multiple transfers share a path.

\mypar{Memory}
The simulator checks per-site feasibility against parameter, optimizer, and activation footprints---the source of the OOM entries in \cref{tab:model-scaling-sim}.

\mypar{RAN dynamics}
Each site replays a TRACTOR-derived SM-reservation trace drawn independently at random from the trace pool, so no two sites are artificially synchronized and the 128--512-site configurations exercise heterogeneous, uncorrelated capacity dynamics.

\section{Implementation Details}
\label{app:implementation}

\mypar{SM-level partitioning via Green Contexts}
\sys{} partitions the GPU's SM pool using the CUDA Green Context API, which provides driver-level isolation of SM subsets.
To keep SM reservation changes off the RAN-slot critical path, each worker pre-creates Green Context streams for all valid SM partitions during initialization.
\sys{} maintains up to 16 such streams, each corresponding to a configurable fraction of the GPU's SMs and consuming approximately 20\,MB of GPU memory, for a total overhead of up to 320\,MB.
The associated tile buffers, stream state, and $(\text{tile size}, \text{spare-SM budget}) \rightarrow \text{completion time}$ profiles are also prepared in advance.
At each slot boundary, the runtime only rebinds the next GEMM tile to the appropriate pre-created context, thereby enforcing the slot-level SM cap while reserving the remaining SMs for training.
This design avoids on-demand resource allocation and per-decision profiling from the admission path, reducing context-switching overhead to below $10\,\mu$s (see \autoref{tab:gpu-sharing}).

\mypar{Asynchronous dispatch and persistent kernels}
Each Green Context is paired with a \emph{persistent kernel} whose lifetime matches the context, backed by a task queue in GPU global memory.
The host enqueues task descriptors (GEMM-tile coordinates, data pointers, output addresses) via mapped zero-copy memory, bypassing the kernel-launch path; a dedicated warp polls the queue head with atomic reads while the remaining warps execute the current tile.
\sys{} intercepts memory-copy completion events on the RAN data path and signals training dispatch via CUDA IPC events, so training GEMMs issue only after RAN decoding data is resident in GPU memory.

\mypar{Memory overhead and tiering}
Beyond the pre-created context streams accounted in \S\ref{sec:eval-overheads}, \sys{} manages GPU HBM and host DRAM as a unified dual-tier pool: the RAN working set occupies a pinned, guaranteed HBM partition; training receives an elastic HBM allocation backed by staging buffers for transparent migration to host memory.
Unlike host-offloading schemes that optimize training throughput in isolation (e.g., ZeRO-Offload~\cite{ren2021zero}), \sys{} applies a \emph{RAN-priority eviction} policy: when PHY memory demand spikes, it demotes training tensors to host memory immediately, preserving RAN latency guarantees at the cost of temporarily reduced training bandwidth.

\mypar{CPU offloading of non-GEMM operators}
\sys{} offloads non-GEMM training operations---layer normalization, element-wise activation functions, and GEMV---to idle host CPU cores.
Because RAN PHY processing is GPU-bound, the host CPU carries mostly lightweight control-plane signaling and has substantial idle capacity.
An adaptive routing policy tracks per-pool queue depth and a sliding-window average of completion times, and assigns each operation to the less-loaded resource at submission time.
This converts idle CPU cycles into additional GPU headroom for the matrix-intensive passes that dominate training compute; the resulting host CPU utilization is accounted in \S\ref{sec:eval-overheads}.

\section{Discussion on Practical Concerns}
\label{app:practical-concerns}

\mypar{General applicability}
While we demonstrate \sys{} with LDPC decoding -- the dominant compute bottleneck for RAN processing -- offloaded to the GPU, the key techniques remain applicable when the entire L1 stack is offloaded (that is supported by NVIDIA Aerial~\cite{nvidia-aerial}); we will explore this once the corresponding hardware is available to us.
Beyond FM training, we will also intend to extend \sys{}'s co-location framework to other non-RAN workloads such as ML inference and RAN digital twins.

\mypar{Coordinator capacity and scaling}
Assuming a widely available datacenter configuration of 200\,Gbps networking and 128 CPU cores for the CPU-only coordinator (e.g., AWS M6in instances), and typical per-site backhaul on the order of hundreds of Mbps~\cite{speedtest}, a single coordinator sustains roughly 1{,}000--2{,}000 sites.
At larger scales, \sys{} shards data and replicates model parameters across multiple coordinators---distributing both bandwidth and computation---following fault-tolerant coordination systems such as Beldi~\cite{beldi}.

\mypar{Robustness to result corruption}
For third-party or multi-tenant deployments, \sys{} can verify distributed GEMM results with random-projection checks~\cite{DBLP:journals/csur/MotwaniR96}: for $C=AB$, sampling random vectors $r,s$ and testing $r^\top(AB)s=(Ar)^\top(Bs)$ detects even single-entry corruption with probability $1-\mathcal{O}(2^{-n})$ at $\mathcal{O}(n)$ cost per check; because the check reduces to GEMVs, it runs in real time on host CPUs~\cite{moe-gen}.

\mypar{Energy considerations}
Harvesting already-provisioned, already-powered RAN GPUs avoids the embodied and provisioning cost of dedicated training clusters.
Decentralized training on such resources can also improve net energy efficiency relative to sub-linearly scaling centralized clusters~\cite{edgetrain,alpa,megatron,deepspeed}.
At our traffic volumes, communication-side energy is negligible relative to compute energy.
Sustained training nonetheless remains bounded by site-specific thermal limits under operator-provisioned power and cooling budgets, and warrants a dedicated energy study~\cite{DBLP:conf/nsdi/KaliaLXFRY25}.

\mypar{Limitations}
(i)~\emph{Convergence scope}: gradient semantics are equivalent to synchronous SGD (Appendix~\ref{app:air-proofs}) and a supporting convergence trace is provided (Appendix~\ref{app:stability-convergence}), but multi-week convergence campaigns on production-scale deployments remain future work.
(ii)~\emph{Single-operator assumption}: \sys{} assumes one operator controls both the RAN and the training workload within an operator-managed execution environment, where users provide models and data while the operator controls the training software~\cite{openai_model_optimization}; stronger isolation for third-party-controlled multi-tenant workloads is future work, and in multi-operator or neutral-host settings \sys{} reduces to passive sharing via Green Context partitioning alone.
(iii)~\emph{Vendor specificity}: SM-level partitioning relies on NVIDIA's Green Context API; porting to accelerators without an equivalent sub-GPU partitioning interface would fall back on weaker software-level isolation.

\end{document}